\documentclass[10pt,preprint]{aastex}

\begin{document}

\shortauthors{Luhman}
\shorttitle{Brown Dwarfs and IMF in Taurus}

\title{New Brown Dwarfs and an Updated Initial Mass Function in Taurus\altaffilmark{1}}

\author{K. L. Luhman}
\affil{Harvard-Smithsonian Center for Astrophysics, 60 Garden Street,
Cambridge, MA 02138}

\email{kluhman@cfa.harvard.edu}

\altaffiltext{1}
{Based on observations performed at Las Campanas, MMT, and Whipple 
Observatories. The MMT Observatory is a joint
facility of the Smithsonian Institution and the University of Arizona.
This publication makes use of data products from the Two Micron All
Sky Survey, which is a joint project of the University of Massachusetts
and the Infrared Processing and Analysis Center/California Institute
of Technology, funded by the National Aeronautics and Space
Administration and the National Science Foundation.}

\begin{abstract}

By combining infrared photometry from the Two-Micron All-Sky Survey
with new optical imaging and spectroscopy, I have performed a
search for young low-mass stars and brown dwarfs in two regions encompassing
a total area of 4~deg$^2$ in the Taurus star-forming region ($\tau\sim1$~Myr).
From this work, I have discovered 15 new members of Taurus. 
In addition, I present seven new members outside of these areas
from the initial stage of a survey of all of Taurus.
These 22 objects exhibit spectral types of M4.5-M9.25 and masses of
0.3-0.015~$M_\odot$ according to the theoretical evolutionary models of 
Baraffe and Chabrier, seven of which are likely to be brown dwarfs. 
Emission in H$\alpha$, He~I, Ca~II, [O~I], and [S~II] and excess emission
in optical and near-infrared bands among some of these objects suggest the 
presence of accretion, outflows, and circumstellar disks. These results add to 
the body of work -- initiated by the first detections of brown dwarf disks by 
Comer\'{o}n and coworkers in 1998 and Luhman in 1999 -- indicating that disks 
around young brown dwarfs are relatively common.
The results from the 4~deg$^2$ survey have been combined 
with previous studies of Taurus to arrive at an initial mass function for a 
total area of 12.4~deg$^2$.
As in the previous IMFs for Taurus, the updated IMF peaks at 
a higher mass (0.8~$M_\odot$) than the mass functions in IC~348 and Orion
(0.1-0.2~$M_\odot$). Meanwhile, the deficit of brown
dwarfs in Taurus appears to be less significant ($\times1.4$-1.8) than 
found in earlier studies ($\times2$) because of a slightly higher brown dwarf 
fraction in the new IMF for Taurus and a lower brown dwarf fraction in the new 
spectroscopic IMF for the Trapezium from Slesnick and coworkers.
The spatial distribution of the low-mass stars 
and brown dwarfs discovered in the two new survey areas closely matches that of 
the more massive members. Thus, on the degree size scales ($\sim3$~pc) probed 
to date, there is no indication that brown dwarfs form through ejection. 

\end{abstract}


\section{Introduction}
\label{sec:intro}

The Taurus star-forming region is an important target for searches for young
low-mass stars and brown dwarfs. Relative to most known star-forming regions,
Taurus exhibits low gas and stellar density, providing
an opportunity to test for variations in the low-mass initial mass function 
(IMF) with star-forming conditions. 
In addition, as one of the nearest major star-forming regions (140~pc), the 
substellar members of Taurus are among the brightest young brown dwarfs in the 
sky, and thus are prime targets for detailed studies of the birth of brown 
dwarfs.

Initial success in discovering members of Taurus down to the hydrogen
burning limit was made through deep imaging of one of its most compact stellar
aggregates, L1495E, at X-ray, optical, and near-infrared (IR) wavelengths
\citep{ss94,lr98}. Subsequently, wide-field optical imaging has proven 
effective in uncovering members at substellar masses and across wide 
areas of the region \citep{bri98,mar01}, particularly when combined with
near-IR photometry from the Two-Micron All-Sky Survey (2MASS) 
\citep{luh00a,bri02,luh03a}. 
Additional brown dwarfs in Taurus have been
identified through observations of the multiple system GG~Tau 
\citep{whi99} and continuum sources \citep{wb03}. 
From this work, 15 Taurus members have been found with spectral types later
than M6, which correspond to substellar masses on an Hertzsprung-Russell (H-R) 
diagram with the evolutionary models of \citet{bar98} and \citet{cha00}.

These surveys have produced the first constraints on the low-mass IMF in
Taurus. \citet{luh00a} derived an IMF for an area of 0.7~deg$^2$ toward four 
of the densest aggregates that was characterized by a peak at 0.8~$M_\odot$ 
and a deficit of brown dwarfs relative to the Trapezium Cluster in Orion 
\citep{luh00b,hc00,mue02}. \citet{bri02} and \citet{luh03a} surveyed an 
additional 7.7~deg$^2$ and for the combined area of 8.4~deg$^2$ measured an 
IMF in which the ratio of the number of brown dwarfs to the number of stars
was again lower than in the Trapezium, this time by a factor of two.
\citet{luh03b} found that this brown dwarf fraction for Taurus did agree 
with the value in the young cluster IC~348, while the peak mass differed 
between Taurus (0.8~$M_\odot$) and IC~348 (0.1-0.2~$M_\odot$). 
The distinctive shapes of the IMFs in Taurus and IC~348 were reflected in the
distributions of spectral types, which peaked at K7 and M5, respectively,
representing unambiguous, model-independent evidence for a significant 
variation of the IMF.
Through recent spectroscopy of a large number of brown dwarf candidates in the
Trapezium Cluster, \citet{sle04} have identified a substantial population of 
faint sources that have relatively early spectral types indicative of stars 
rather than brown dwarfs.  They suggested that these
contaminating sources have resulted in overestimates of the brown dwarf
fraction in the Trapezium in previous studies of the cluster, which were 
based almost exclusively on photometry. After correcting for these contaminants,
\citet{sle04} measured a brown dwarf fraction for the Trapezium that was a
factor of $\sim1.8$ greater than that in Taurus. 
The theoretical implications of these IMF measurements in Taurus
have been investigated in these surveys \citep{luh00a,bri02,luh03b} and in 
subsequent work \citep{kb03b,del04,goo04}.

To continue to improve the number statistics of the IMF measurements in 
Taurus, to sample a larger fraction of the region, and to find additional 
targets for studies of brown dwarf formation, I have surveyed two 
new regions in Taurus encompassing a total area of 4~deg$^2$. In this paper,
I describe $I$ and $z\arcmin$ imaging of these regions (\S~\ref{sec:phot})
and optical spectroscopy of candidate members (\S~\ref{sec:spec}), which are 
selected from a combination of this optical photometry and $JHK_s$ data 
from 2MASS (\S~\ref{sec:cand}). I then measure the spectral types of these
candidates and determine their status as field stars or Taurus members 
(\S~\ref{sec:class}), discuss the physical properties of the confirmed 
members (\S~\ref{sec:prop}), and evaluate the completeness of the survey 
(\S~\ref{sec:comp}). 
For all known Taurus members in the new survey area, I estimate extinctions, 
luminosities, and effective temperatures, and construct an H-R diagram
(\S~\ref{sec:hr}).  Using the theoretical evolutionary models, I infer 
individual masses for these sources and derive an IMF from the combination 
of the previously studied 8.4~deg$^2$ regions (Field 1) and these new 
4~deg$^2$ regions (Field 2) (\S~\ref{sec:imf}). 
Finally, I examine the implications of this new
measurement of the IMF in Taurus for recent theoretical predictions 
of the IMF of stars and brown dwarfs (\S~\ref{sec:imfcompare}).

\section{Observations}
\label{sec:obs}

\begin{figure}
\plotone{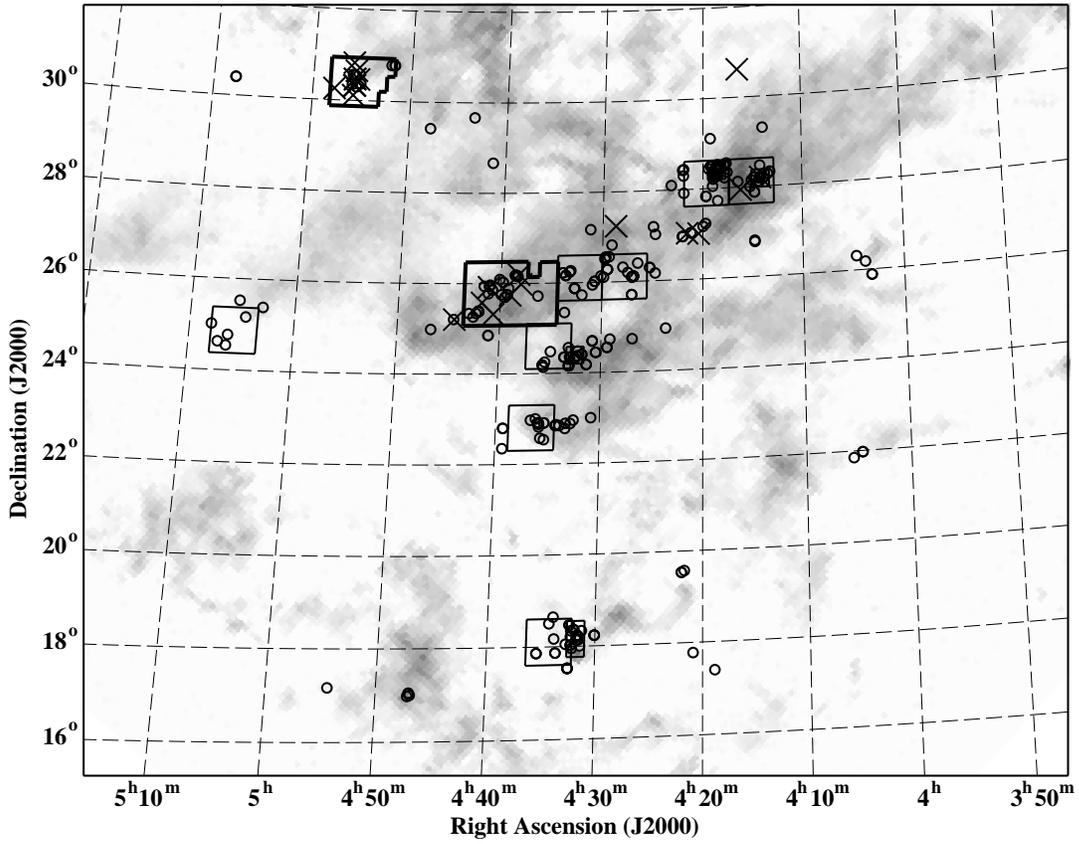}
\caption{
Spatial distribution of the previously known members of the Taurus 
star-forming region ({\it circles}) and the 22 new members found in this work
({\it crosses}) shown with emission in $^{12}$CO ({\it grayscale}, T. Megeath,
private communication).
The regions surveyed for new members by \citet{bri98}, \citet{luh00a},
\citet{bri02}, and \citet{luh03a} ({\it thin lines}) and in this 
work ({\it thick lines}) are referenced henceforth as Fields 1 and 2,
respectively.
}
\label{fig:map1}
\end{figure}

The regions in Taurus previously imaged at $I$ and $z\arcmin$ by
\citet{bri98}, \citet{luh00a}, \citet{bri02}, and \citet{luh03a} are
indicated in Figures~\ref{fig:map1} and \ref{fig:map2}, which encompass
a total area of 8.4~deg$^2$. 
For the new imaging in this study, I selected two areas toward known
groups of young stars, which are shown in those maps. The north and south 
regions are centered at 
$\alpha=4^{\rm h}55^{\rm m}00^{\rm s}$, $\delta=30\arcdeg24\arcmin30\arcsec$ and
$\alpha=4^{\rm h}39^{\rm m}00^{\rm s}$, $\delta=25\arcdeg46\arcmin00\arcsec$
(J2000) and cover 1.32 and 2.84~deg$^2$, respectively. Henceforth in this work, 
I refer to these regions in the old and new surveys as Field 1 and Field 2.
In addition, I will present spectra from an initial sample of candidates
drawn from a survey across 225~deg$^2$ encompassing all of the Taurus 
star-forming region, which will be referenced as the 225~deg$^2$ survey.

\subsection{Photometry}
\label{sec:phot}

Images of the two areas in Field 2 were obtained with the four shooter camera 
at the Fred Lawrence Whipple Observatory 1.2~m telescope on 2002 December 6-11.
The instrument contained four $2048\times2048$
CCDs separated by $\sim45\arcsec$ and arranged in a $2\times2$ grid. After
binning $2\times2$ during readout, the plate scale was $0\farcs67$~pixel$^{-1}$.
The fields were divided into grids in right
ascension and declination where the grid cells were separated by $20\arcmin$.
At each cell, two positions separated by $1\arcmin$ in right
ascension and declination were observed.
At each position, images were obtained with exposure
times of 30 and 900~s at $I$ and 20 and 600~s at $z\arcmin$.
The images were bias subtracted, divided by dome flats, registered, and
combined into one image at each band and exposure time.
During one of the photometric nights, additional short exposures were obtained
in a grid of pointings separated by $40\arcmin$. These exposures overlapped
with all of the pointings in the full maps and thus enabled checking of the
photometric calibration. 

Photometry and image coordinates
of the sources in these data were measured with DAOFIND and
PHOT under the IRAF package APPHOT.
Aperture photometry was extracted with a radius of five pixels. The background
level was measured in an annulus around each source and subtracted from
the photometry, where the inner radius of the annulus was five pixels and the
width was one pixel.
The photometry was calibrated in the Cousins $I$ system by combining
data for standards across a range of colors \citep{lan92} with the
appropriate aperture and airmass corrections. The $z\arcmin$ data were 
calibrated by assuming $I-z\arcmin=0$ for A-type standard stars.
This type of calibration at $z\arcmin$ is sufficient since the analysis in 
this study relies only on the relative precision of the $z\arcmin$ photometry.
To compare this $z\arcmin$ photometry to data from another instrument, one 
would need to calibrate the latter in the manner applied to these data and to 
account for any differences in the filter and instrument transmission profiles. 
Saturation in the short exposures occurred near magnitudes of 11-12 in both 
filters. The completeness limits of the long exposures were $I\sim19$ and 
$z\arcmin\sim18$. The plate solution was derived from
coordinates measured in the 2MASS Point Source Catalog for stars that 
appeared in the optical images and were not saturated.
During the course this data reduction, I discovered an arithmetic error in
the photometric calibration of the data from \citet{luh00a} such that those 
$z\arcmin$ measurements were too bright by 0.55~magnitudes. 
This error had no affect on any of
the analysis in that study since the $z\arcmin$ photometry was used only
for computing $I-z\arcmin$, and only the relative values of these colors
were relevant.

\subsection{Spectroscopy}
\label{sec:spec}

I performed low-resolution optical spectroscopy on 49 candidate members
of Taurus.
The selection of these sources is described in \S~\ref{sec:cand}. 
Table~\ref{tab:log} summarizes the observing runs and instrument configurations
for these data. In Tables~\ref{tab:field} and \ref{tab:new}, I indicate the
night on which each star was observed.
On 2003 December 14, I also obtained a spectrum of Haro~6-32, which is a 
previously known member of Taurus within the areas surveyed in this 
study and lacking a published spectral type. The spectroscopy of this 
sample was performed with the Inamori Magellan Areal Camera and 
Spectrograph (IMACS) on the Magellan~I telescope at Las Campanas Observatory
and the Red Channel spectrograph at the MMT Observatory. 
Each spectrum was collected with the slit rotated to the parallactic angle. 
The exposure times ranged from 120 to 1800~s. 
After bias subtraction and flat-fielding,
the spectra were extracted and calibrated in wavelength with arc lamp data.
The spectra were then corrected for the sensitivity functions of the detectors,
which were measured from observations of spectrophotometric standard stars.
For some of the objects observed with Red Channel on the MMT, noticeable 
fringing remains in the final spectra at long wavelengths, as seen in
Figures~\ref{fig:spec1} and \ref{fig:spec2}. The spectra with IMACS on 
Magellan exhibit a gap in wavelength coverage at 6640-6740~\AA\ that
is produced by the boundary between two adjacent CCDs.

\section{New Members of Taurus}
\label{sec:new}

\subsection{Selection of Candidate Members}
\label{sec:cand}

A nearby population of young stars and brown dwarfs will exhibit colors and 
absolute photometry that are distinct from those of most field stars. 
Therefore, color-magnitude diagrams composed of optical and near-IR photometry 
are useful tools for distinguishing candidate members of a young cluster from 
field stars. However, the presence of variable extinction increases the 
contamination of background stars in the regions of these diagrams 
inhabited by members and prevents reliable mass estimates for candidate 
members because reddened high-mass members and unreddened low-mass members
can have the same optical color and magnitude.
To better separate young objects from field stars and to improve the 
ability to estimate masses of candidates in such diagrams, the photometry 
in these diagrams can be corrected for reddening. 
Therefore, I have constructed extinction-corrected diagrams of
$I-K_s$ versus $H$ and $I-z\arcmin$ versus $H$ from the optical photometry 
in this work and near-IR data from the 2MASS Point Source Catalog, where
the reddenings toward individual stars are estimated in the manner described 
by \citet{luh04}. I then use these diagrams to identify candidate young 
stars and brown dwarfs in the Taurus fields surveyed in this work.
This dereddening method requires no knowledge of spectral types or membership
and can be applied uniformly to photometry to better separate cluster
members from field stars in a color-magnitude diagram. More accurate
extinctions for the confirmed cluster members will be estimated in
\S~\ref{sec:hr} by incorporating the spectral type measurements.

When the known members of Taurus are placed on a H-R diagram, most of them
appear above the model isochrone for an age of 10~Myr from \citet{bar98}
(\S~\ref{sec:hr}). In the diagram of $I-K_s$ versus $H$ in
Figure~\ref{fig:4sh2}, I plot the 10~Myr isochrone from \citet{bar98} for 
masses of 0.015 to 1~$M_{\odot}$ at a distance modulus of 5.76 \citep{wic98}.
This isochrone was converted to photometric magnitudes from predicted
effective temperatures and bolometric luminosities in the
manner described by \citet{luh03a}. I have defined a
boundary below this isochrone to separate candidate members of Taurus and
probable field stars, as shown in Figure~\ref{fig:4sh2}.
For $I-z\arcmin$ versus $H$, I defined a similar boundary below the lower
envelope of the sequence of known members in the fields observed by 
\citet{luh00a}. 
I am primarily interested in identifying candidates at lower reddenings
that could fall within the extinction limit that will be used to define
the IMF sample for this survey (\S~\ref{sec:imf}). As a result, 
I consider only sources with $A_V\leq8$ in Figure~\ref{fig:4sh2}.
For spectroscopy, I selected the 28 objects that appeared above both of these
boundaries, were not previously known members, and had dereddened $I-K_s>2.5$ 
($\gtrsim$M4) and accurate photometry ($I<19$). Sources with $I>19$ cannot
be reliably identified as field stars or candidate members based these
color-magnitude diagrams because of the photometric uncertainties. 
Nevertheless, I did include one object just beyond this magnitude limit which 
appeared well to the right of the boundaries in Figure~\ref{fig:4sh2} and 
thus was a promising candidate. The impact of the unobserved candidates 
at $I-K_s\leq2.5$ and the sources with $I>19$ on the completeness of this
survey is discussed in \S~\ref{sec:comp}.
Two of these objects in my spectroscopic sample were previously identified
as candidate low-mass members of Taurus through detections of $K$-band excess
emission. Those near-IR data were collected by \citet{itg96} and the astrometry
and photometry for the two candidates, ITG~2 and 34, were reported by 
\citet{itn99}.

The boundaries used in separating candidate members from probable field stars
in Figure~\ref{fig:4sh2} were also applied to the data from \citet{bri02},
after shifting the $I-z\arcmin$ boundary to match the color system in that work.
In this way, I identified four remaining candidates from that study that lacked
spectroscopy, all of which were near the selection boundaries and thus were
marginal candidates. These four stars were included in my spectroscopic sample.
Finally, I have recently begun a search for young low-mass star and brown 
dwarfs across 225~deg$^2$ encompassing all of the Taurus star-forming region. 
These candidates were selected with data from the USNO-B1 and Two-Micron
All-Sky Survey (2MASS) catalogs. The selection criteria for these candidates 
will be described in a future work when the survey is completed. 
I include here the spectroscopy of the first 17 candidates from that survey.
These sources can be identified in Tables~\ref{tab:field} and \ref{tab:new}
by their lack of $I$ photometry.
One of these objects, IRAS~04414+2506, was previously identified as a 
candidate member of Taurus by \citet{ken94} on the basis of its 
mid-IR excess emission.

\subsection{Classification of Candidate Members}
\label{sec:class}

I measure spectral types and assess membership for the 49 candidate Taurus
members in my spectroscopic sample by applying the methods of
classification described in my previous studies of Taurus and other young 
populations \citep{luh99,luh04,luh03a,luh03b}. 
In this way, I classify 27 candidates as field stars.
Astrometry and photometry for these sources are listed in Table~\ref{tab:field}.
The spectra for the 22 remaining candidates are shown in 
Figures~\ref{fig:spec1} and \ref{fig:spec2}. These data exhibit evidence of 
youth, and thus membership in Taurus, in the form of the weak K~I and Na~I
absorption features that are characteristic of pre-main-sequence objects
\citep{mar96,luh99}, as illustrated in Figure~\ref{fig:nak} in a comparison
of a young star and a field dwarf from my sample.
These objects are clearly not field giants according to various spectral 
features in these data, such as CaH \citep{kir97}.
The youth of several of these sources is further indicated
by the presence of strong emission in H$\alpha$ that is
above the levels observed for active field dwarfs \citep{giz00,giz02}.
In addition, the membership of five objects is independently established by
the presence of reddening in their spectra and their positions above the main
sequence for the distance of Taurus, which indicate that they cannot be field
dwarfs in the foreground or the background of the cloud, respectively.
Astrometry, photometry, spectral types, and evidence of membership for these
22 new members of Taurus are provided in Table~\ref{tab:new}.
For a few sources that exhibited evidence of veiling in their spectra, 
such as 2MASS~J04414825+2534304, J04411078+2555116, and
J04141188+2811535, the spectral types are based primarily on the reddest 
absorption bands ($>$8000~\AA) where the excess continuum emission should
be weakest.

The membership and youth of the 22 new sources presented 
in this study are conclusively established by the criteria listed in 
Table~\ref{tab:new}. Future measurements of additional diagnostics such
as Li are not required for confirmation of membership since
that is already determined in this work. This also applies to many of the
other young late-type objects discovered in previous surveys of star-forming
regions. For instance, \citet{bar04a} allegedly ``confirmed" the youth and
membership of LS-RCrA~1 through measurements of Li, gravity-sensitive K~I,
and a radial velocity. However, the youth of this source was not
in question given the intensity of the H$\alpha$ emission and other
permitted lines, presence of emission in H$_2$ and forbidden transitions, 
and strength of the gravity-sensitive Na~I and K~I in the discovery data 
from \citet{fer01}. Similarly, \citet{bar04b} obtained high-resolution spectra
of three sources discovered by \citet{bri02}. Although multiple lines of 
evidence were used to firmly establish the youth and membership of these 
objects by \citet{bri02}, \citet{bar04b} referred to them as
``candidate" members and proceeded to ``confirm" their membership through the 
detection of Li absorption. Indeed, \citet{bar04b} failed to explain in any 
way why the evidence of membership presented by \citet{bri02} was insufficient.
These claims of ``confirmation" by \citet{bar04a} and \citet{bar04b} 
are misleading and overstated since they imply the presence of an 
uncertainty in the nature of the sources in the original discovery papers 
when no such uncertainty existed. 

\subsection{Properties of New Members}
\label{sec:prop}

\subsubsection{Disks, Accretion, and Outflows}
\label{sec:disks}

In this section, I examine the available data for the 22 new members of Taurus 
for signatures of circumstellar disks, accretion, and outflows.
I begin by plotting these sources in the diagram of $H-K_s$ versus $J-H$ 
in Figure~\ref{fig:jhhk} to check for the presence $K$-band excess emission
from circumstellar material.  For comparison, I include
the sequences of dwarfs and giants from \citet{leg92} and \citet{bb88} for
spectral types of $\leq$M9~V and $\leq$M5~III, respectively.
As shown in Figure~\ref{fig:jhhk}, most of the 22 new members have colors
that agree well with reddened values of dwarfs at the spectral types in
question. This result indicates that dwarf colors are good approximations
of the intrinsic colors of young late-type objects, as found previously 
by \citet{luh99} and \citet{bri02}. Meanwhile, a few of these sources do
exhibit significant $K$-band excess ($\gtrsim0.15$~mag), namely 
2MASS~J04381486+2611399 (M7.25), J04141188+2811535 (M6.25), and 
J04161210+2756385 (M4.75). The excess for the first object is particularly
large. Although 2MASS~J04574903+3015195 (M9.25) also appears to have an excess 
in Figure~\ref{fig:jhhk}, I do not consider this excess significant because 
the photometry has large uncertainties at the faint levels of this source.
These detections of K-band excess emission, particularly for the brown dwarf
2MASS~J04381486+2611399, provide further evidence that young brown dwarfs have 
circumstellar disks, as first indicated by the detections of IR excess emission 
from substellar objects by \citet{com98} and \citet{luh99}. 

As seen in Figures~\ref{fig:spec1} and \ref{fig:spec2}, some of the
new Taurus members exhibit H$\alpha$ emission at high levels
($W_{\lambda}>100$~\AA) as well as emission in He~I and Ca~II.
Strong H$\alpha$ emission and the presence of He~I and Ca~II
emission are suggestive of intense accretion when observed in
young stars \citep{muz98,ber01}.
In addition, during the spectral classifications, I found that blue 
continuum veiling appeared to be present in the spectra
of 2MASS~J04414825+2534304 (M7.75), J04411078+2555116 (M5.5), and 
J04141188+2811535 (M6.25), all of which also show very strong emission in 
H$\alpha$. As an example, in Figure~\ref{fig:veil} I compare the spectrum of one
of these sources, 2MASS~J04141188+2811535, to the spectrum of another
Taurus member, 2MASS~J04552333+3027366, with the same spectral type 
as inferred from the VO band strengths at wavelengths beyond 8000~\AA.
This comparison clearly reveals the presence of excess emission that 
increasingly dilutes the photospheric absorption features at shorter
wavelengths, which is direct evidence of intense accretion.
Detailed analysis of the accretion in these objects will be provided
through high-resolution spectroscopy by \citet{muz04}.

In Figure~\ref{fig:lines}, I present detections of emission in [O~I] in
three young brown dwarfs, 2MASS~J04442713+2512164, J04381486+2611399, and
J04414825+2534304. Emission in [S~II] is also found in the first two sources.
Emission in forbidden transitions is typical of Herbig-Haro objects, 
Class~I sources, and some T Tauri stars, where the emission is believed to 
arise predominantly in jets and outflows \citep{ken98}. 
These three new brown dwarfs are the coolest known young objects with 
emission in forbidden lines.
Previous discoveries of forbidden line emission in slightly earlier objects
have been made by \citet{fer01} and \citet{muz03}.

\subsubsection{Spatial Distribution}
\label{sec:spatial}

I now discuss the locations and spatial distribution of the 22 new members
of Taurus discovered in this study. 

Seven of the new members are from the 225~deg$^2$ survey
and thus are scattered across the region, as indicated in Figure~\ref{fig:map1}.
2MASS~J04202555+2700355 (M5.25), 04213459+2701388 (M5.5), and 
04284263+2714039 (M5.25) are between L1495 and L1529. 
2MASS~J04442713+2512164 (M7.25) is just beyond the eastern border of the
southern area in Field 2, while 
2MASS~J04163048+3037053 (M4.5) is $2\arcdeg$ north of L1495.
2MASS~J04141188+2811535 (M6.25) is in a $20\arcmin\times20\arcmin$ area
observed by \citet{bri98} and \citet{luh00a}. 
A $1\arcdeg\times1\arcdeg$ field from \citet{bri02} also contains this
source, as well as 2MASS~J04161210+2756385 (M4.75). These two objects were
not discovered in those surveys because they were just above the saturation
limits of the optical photometry. The implications of those saturation limits
for the completeness of the previous surveys in Taurus are discussed in 
\S~\ref{sec:comp}.

\begin{figure}
\plotone{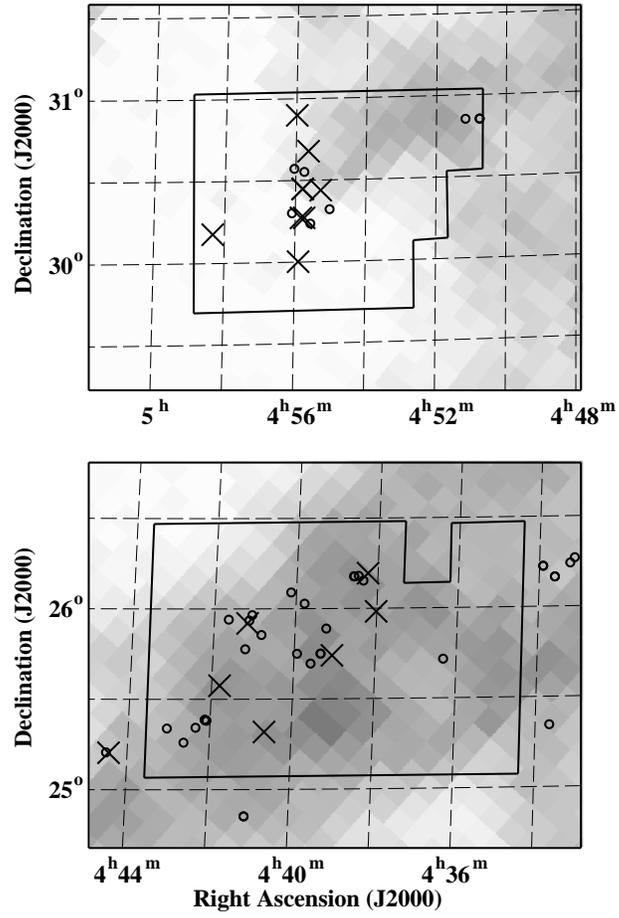}
\caption{
Expanded versions of Figure~\ref{fig:map1} centered on the regions surveyed
in this work (Field 2), which encompass a total area of 4~deg$^2$.
}
\label{fig:map2}
\end{figure}

\begin{figure}
\epsscale{0.8}
\plotone{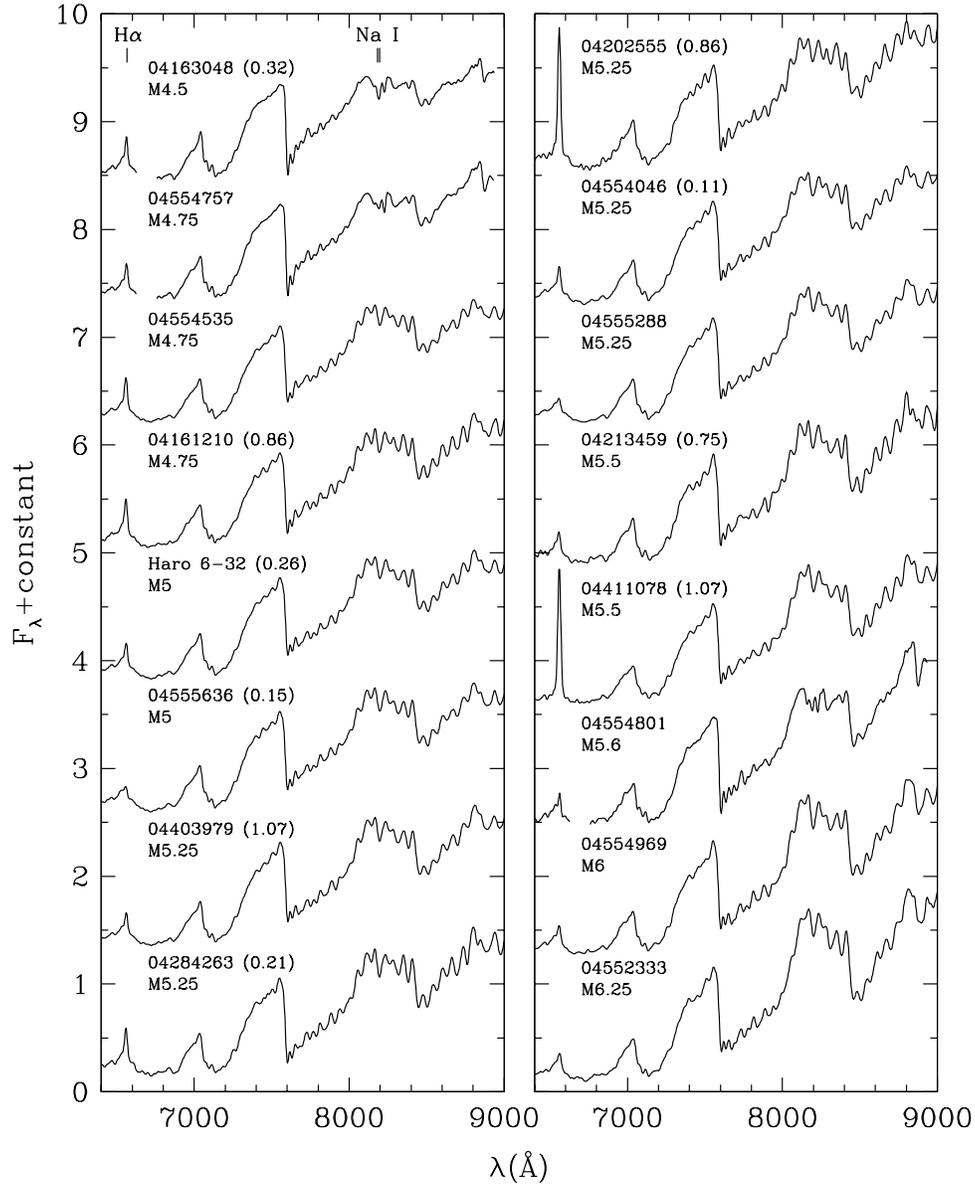}
\caption{
Low-resolution spectra of new members of the Taurus star-forming region.
We also show the spectrum of Haro 6-32, a previously known member within 
the areas surveyed in this study that lacks a published spectral type. 
The spectra have been corrected for extinction, which
is quantified in parentheses by the magnitude difference of the reddening
between 0.6 and 0.9~\micron\ ($E(0.6-0.9)$, \S~\ref{sec:hr}).
The data are displayed at a resolution of 18~\AA\ and are normalized at
7500~\AA.}
\label{fig:spec1}
\end{figure}

\begin{figure}
\epsscale{1}
\plotone{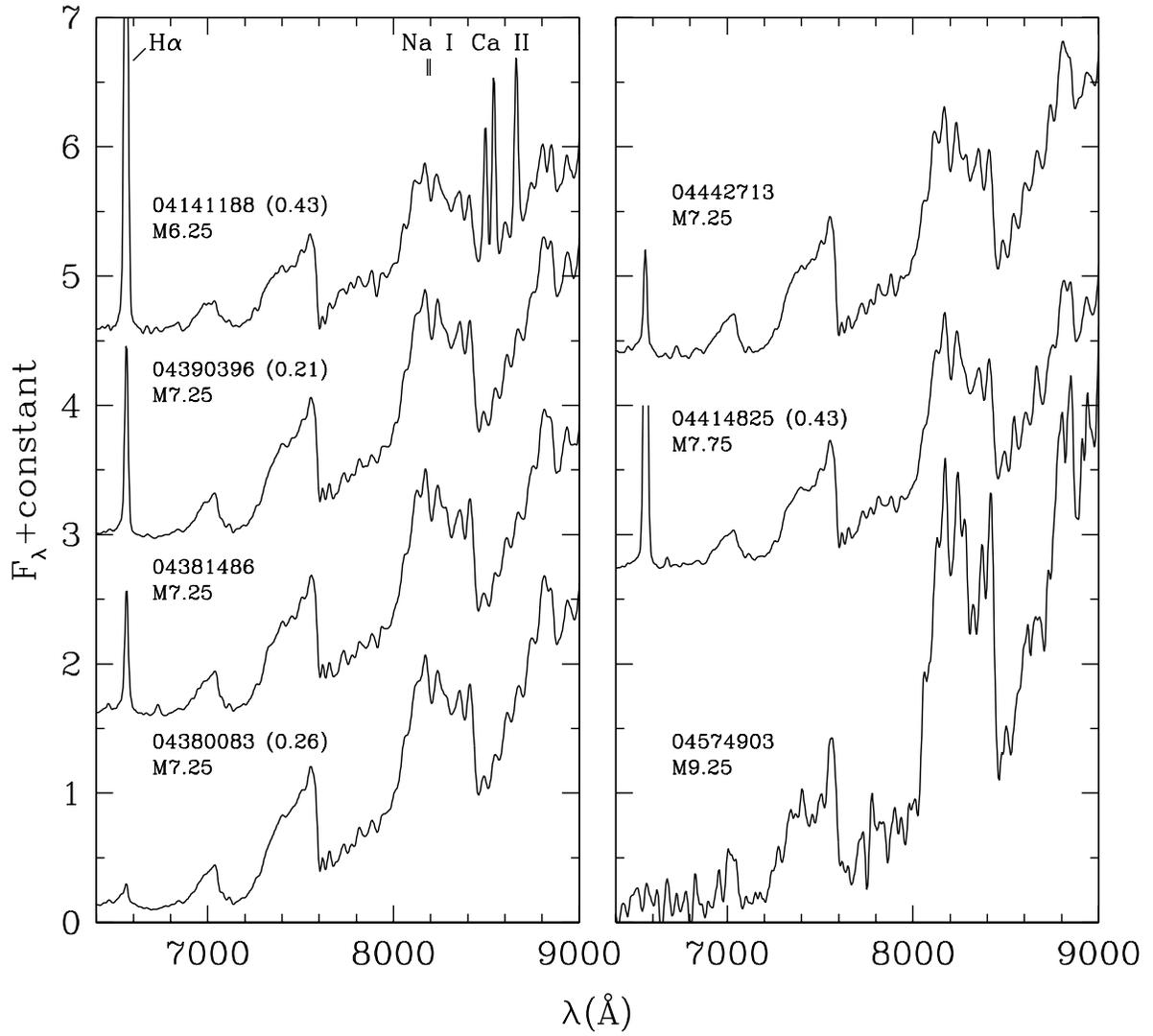}
\caption{
Same as in Figure~\ref{fig:spec1}.}
\label{fig:spec2}
\end{figure}

\begin{figure}
\epsscale{0.75}
\plotone{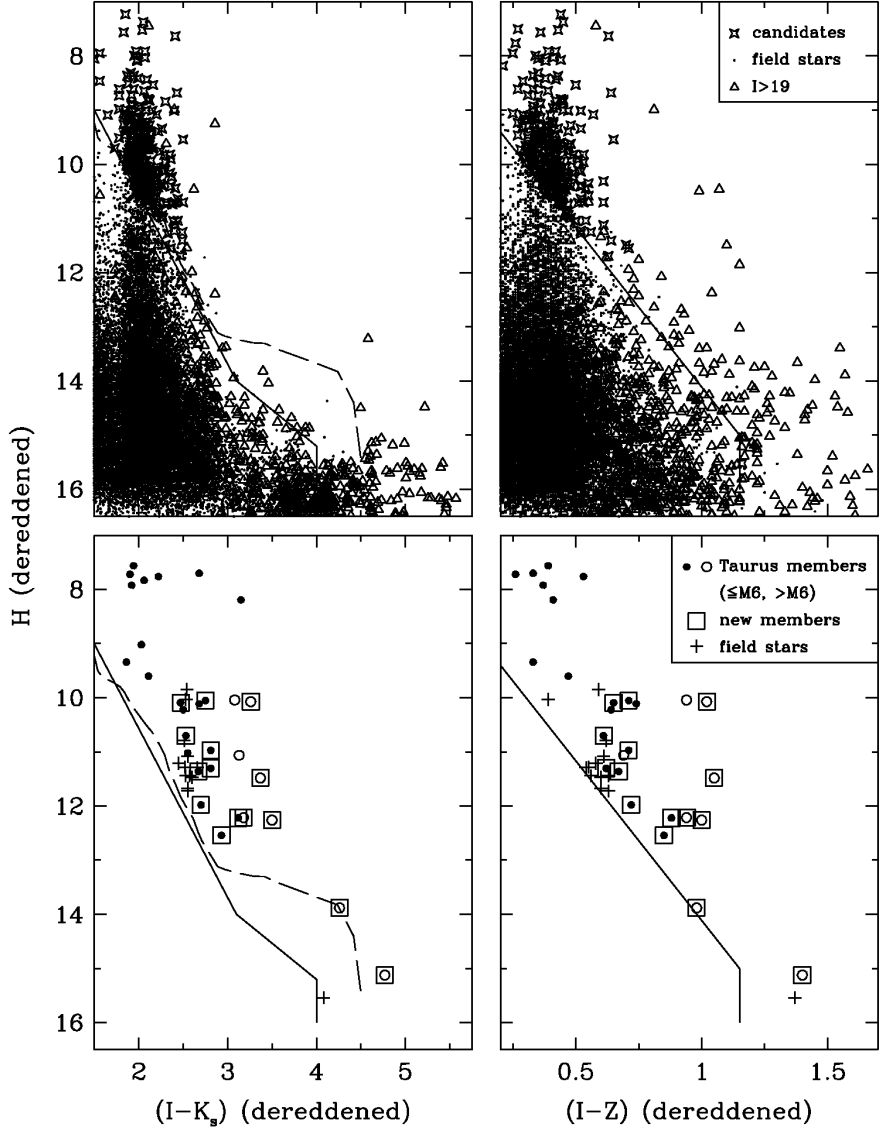}
\caption{
Extinction-corrected color-magnitude diagrams for stars with $A_V\leq8$ in 
Field 2 (Figure~\ref{fig:map2}).
The bottom panels show the stars that have been spectroscopically confirmed as
Taurus members at types $\leq$M6 and $>$M6 ({\it large points and open circles})
and indicate the ones found in this work ({\it boxes}).
Members later than M6 are likely to be brown dwarfs by the H-R diagram and
evolutionary models in Figure~\ref{fig:hr}.
The field stars identified spectroscopically in this study are also shown 
({\it plusses}). The top panels contain the remaining stars, which consist of 
objects that are candidate members by their locations above both solid
boundaries ({\it stars}), objects that are likely to be field stars by their 
locations below either of the solid boundaries ({\it small points}), 
and stars that are too faint to be reliably identified as candidate members or
field stars ($I>19$, {\it triangles}). The dashed line is the 10~Myr isochrone 
(1-0.015~$M_{\odot}$) from the evolutionary models of \citet{bar98}. 
}
\label{fig:4sh2}
\end{figure}

\begin{figure}
\epsscale{1}
\plotone{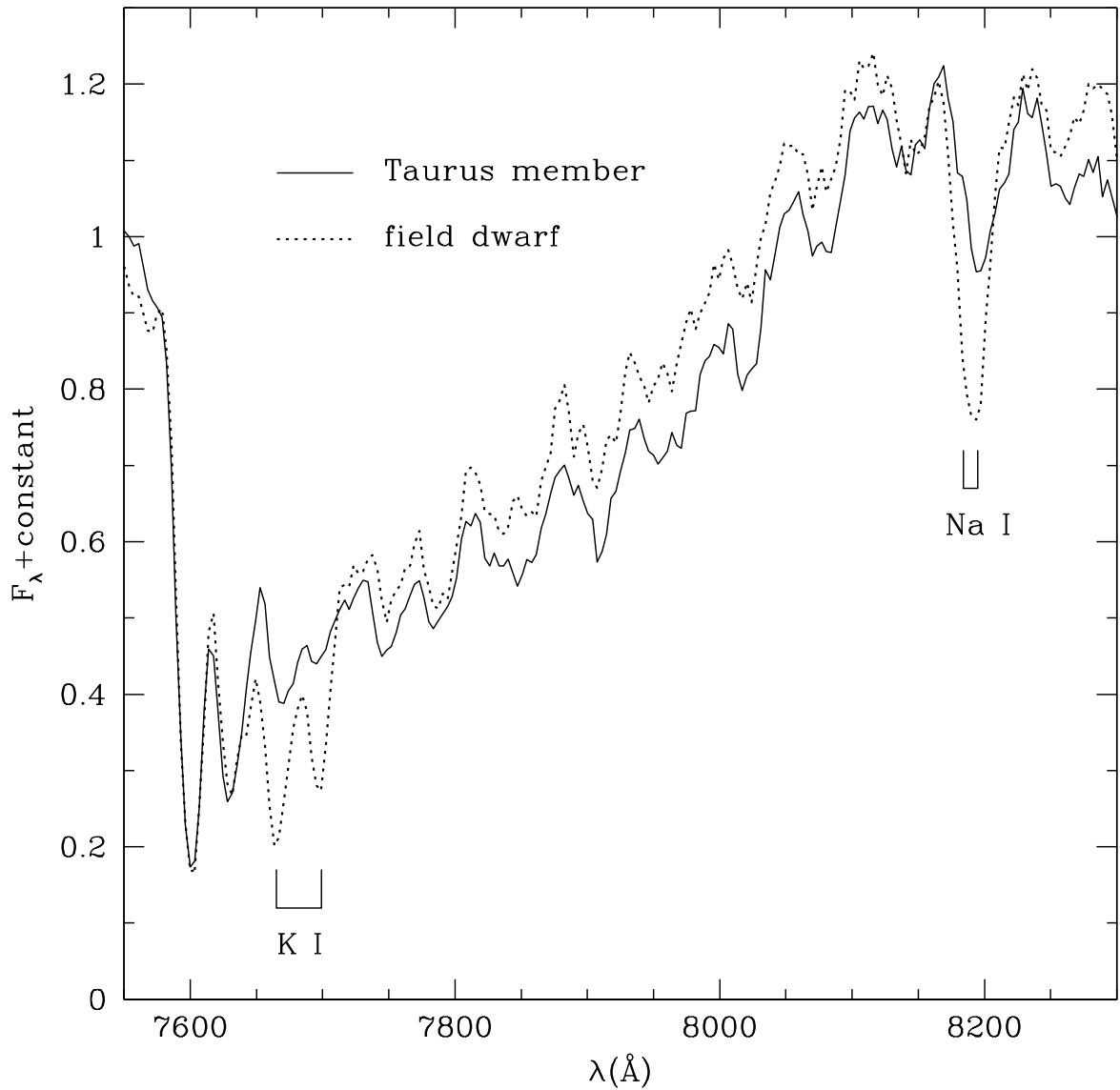}
\caption{
The K~I and Na~I absorption lines in two candidate members of Taurus.
These gravity-sensitive lines are 
strong in 2MASS~J04425686+2610546 ({\it dotted line}) and weak in 
2MASS~J04554535+3019389 ({\it solid line}), indicating that the former is
a field dwarf and the latter is a pre-main-sequence object and thus a
member of Taurus.
The data are displayed at a resolution of 10~\AA\ and are normalized at
7500~\AA.
}
\label{fig:nak}
\end{figure}

\begin{figure}
\plotone{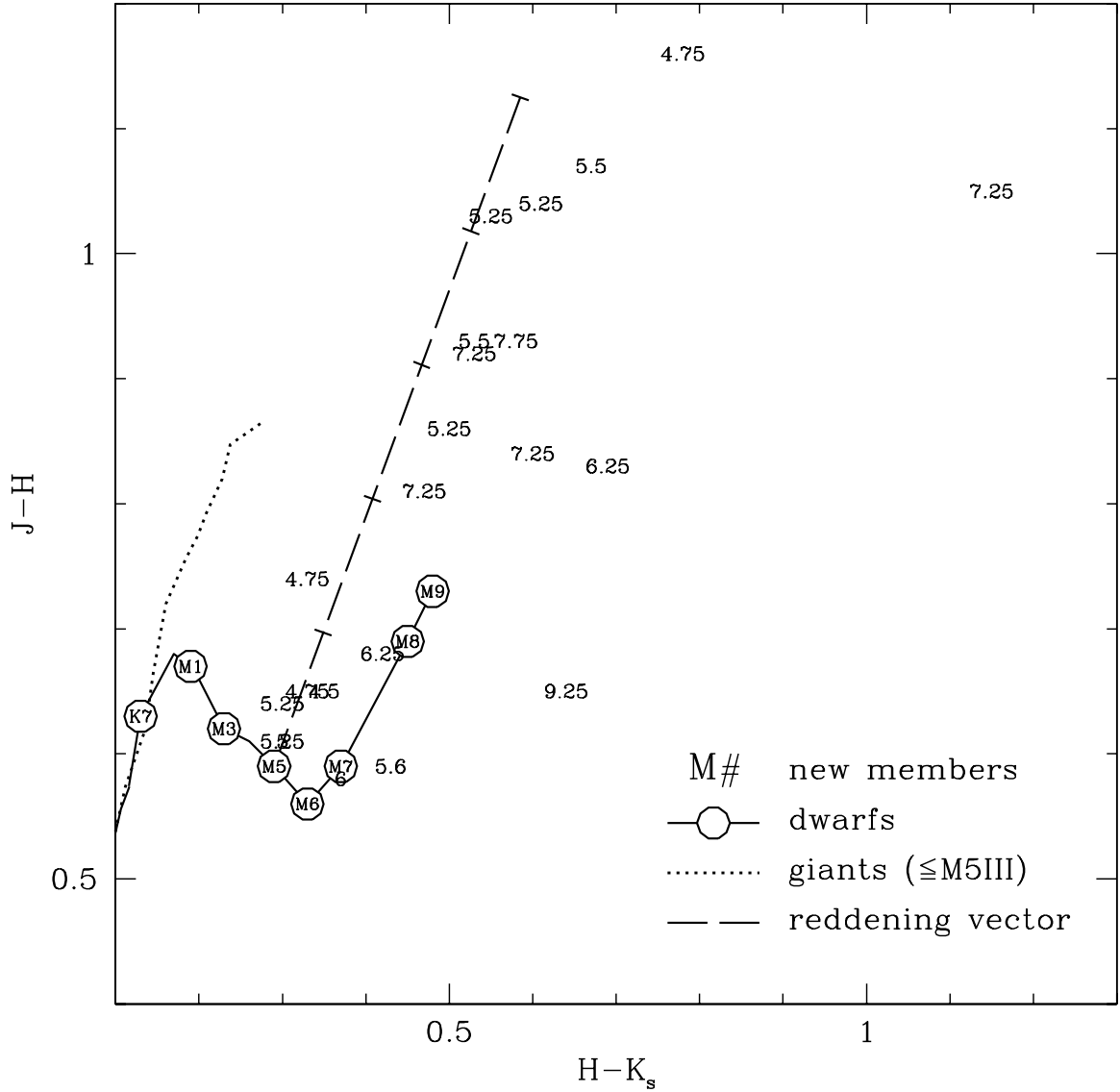}
\caption{
$H-K_s$ versus $J-H$ from 2MASS for the 22 new members of Taurus, which
are represented by the M subclass of their spectral types. 
I include sequences for typical field dwarfs ({\it solid line}; $\leq$M9V)
and giants ({\it dotted line}; $\leq$M5~III) and a reddening vector
originating at M5V with marks at intervals of $A_V=1$ ({\it dashed line}).
Some of these new low-mass stars and brown dwarfs exhibit $K$-band emission
in excess above that expected from reddening, indicating the presence of 
circumstellar material.}
\label{fig:jhhk}
\end{figure}

\begin{figure}
\plotone{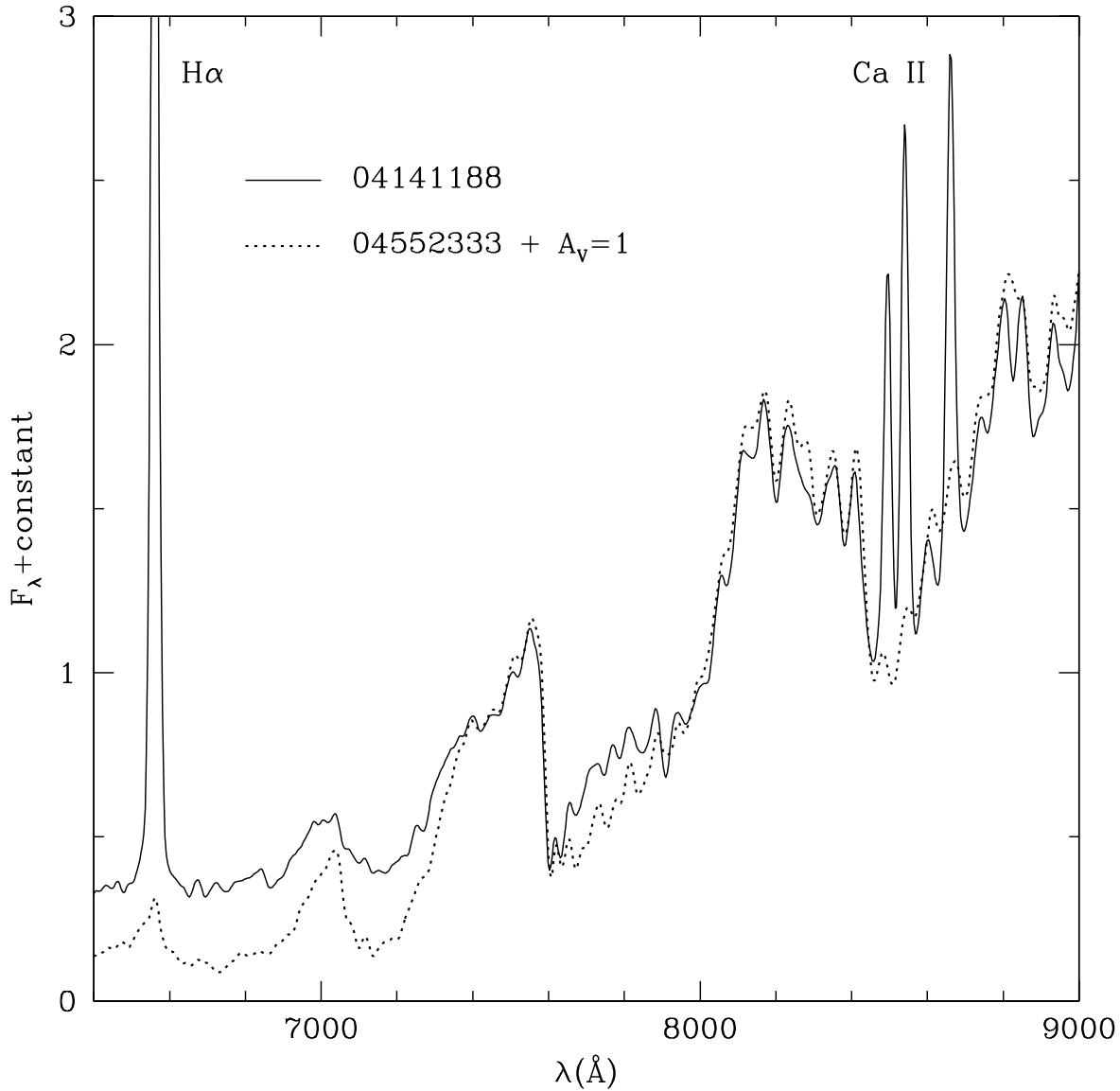}
\caption{
The spectrum of the new Taurus member 2MASS~J04141188+2811535 ({\it solid line})
compared to the data for another new member, 2MASS~J04552333+3027366
({\it dotted line}), at the same spectral type of M6.25 as derived from
the data longward of 8000~\AA. The spectrum of the 
latter has been reddened to match the slope of 2MASS~J04141188+2811535.
Relative to the comparison star, 2MASS~J04141188+2811535
exhibits a strong excess of continuum at short wavelengths, which together
with the strong H$\alpha$ emission ($W_{\lambda}\sim200$~\AA) indicate the 
presence of intense accretion.
These spectra are displayed at a resolution of 18~\AA\ and are normalized at
7500~\AA.}
\label{fig:veil}
\end{figure}

\begin{figure}
\epsscale{0.75}
\plotone{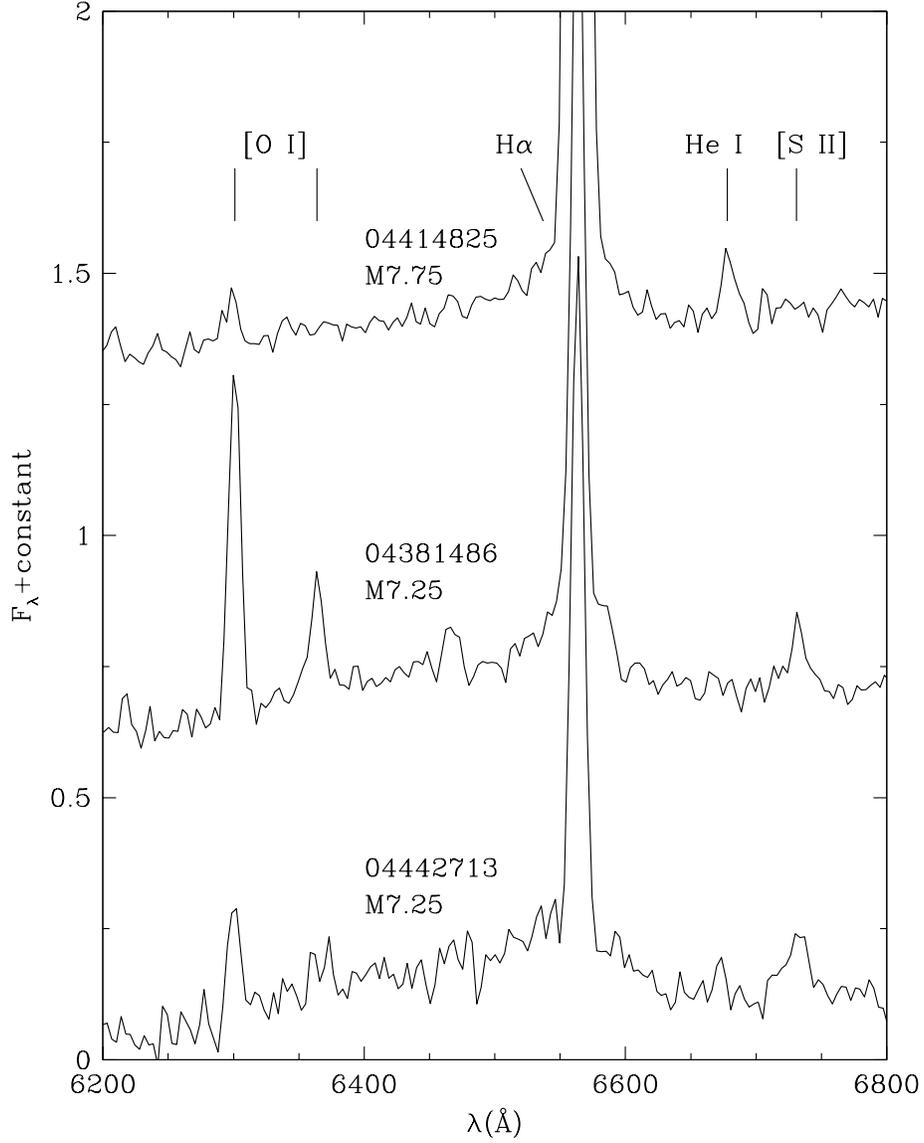}
\caption{
Emission lines in three of the new brown dwarfs in Taurus.
The emission in the forbidden transitions suggests the presence of outflows,
particularly in the lower two sources, while the strength of H$\alpha$ 
($W_{\lambda}\sim250$~\AA) and the detection of He~I in the upper object 
are indicative of accretion.
}
\label{fig:lines}
\end{figure}

The remaining 15 new members are within Field 2.
Figure~\ref{fig:map2} compares the positions of these new members and the 
previously known members, which correspond to the low- and high-mass 
members, respectively.
These results provide useful constraints on theories for the
formation of brown dwarfs. For instance, \citet{rc01}, \citet{bos01}, and 
\citet{bat02} have suggested that brown dwarfs could
form as protostellar sources whose accretion is prematurely halted by
ejection from multiple systems. The resulting spatial distributions 
of the ejected brown dwarfs might differ from those of the stellar members. 
Alternatively, if brown dwarfs and stars form in the same manner,
then they should exhibit similar spatial distributions. \citet{kb03a} 
and \citet{kb03b} have investigated these scenarios in detail through
dynamical modeling of Taurus-like aggregates of stars and brown dwarfs.
In the model in which stars and brown dwarfs form in the same manner,
they found that most of the brown dwarfs indeed shared the same
spatial distributions as the stars and a high binary fraction, while a small 
tail of higher velocity ($v>1$~km~s$^{-1}$) brown dwarfs were distributed
more widely than the stars and were predominantly single.
In comparison, the ejection model also produces widely distributed 
single brown dwarfs, but they are the dominant population among the brown
dwarfs and thus exhibit a shallow density distribution \citep{kb03a}.
Based on their interpretation of measurements of the IMF
in Taurus, the Trapezium, and the field, \citet{kb03b} tended to favor 
the ejection model (\S~\ref{sec:imfcompare}). However, 
\citet{bri02} found no statistically significant differences in the 
distribution of the high- and low-mass members of Taurus.  
Similarly, I find that the spatial distribution of the new low-mass members 
closely matches that of the previously known, more massive young stars, 
as demonstrated in Figure~\ref{fig:map2}. 
Thus, the available data are better 
reproduced by the model in which stars and brown dwarfs have a common
formation mechanism \citep{kb03a} than the ejection model \citep{kb03b}. 
A definitive test of the two models will require measurements of 
the density distribution of young brown dwarfs on larger size scales. 
Nevertheless, from the surveys of Taurus to date, there is no evidence 
for different spatial distributions 
between stars and brown dwarfs.

Finally, I note that two of the new members, 2MASS~J04554801+3028050 (M5.6)
and J04554757+3028077 (M4.75), are separated by only $6\arcsec$. Considering
the very low stellar density in Taurus, these objects are likely to be 
components of a binary system rather than unrelated members.

\subsection{Completeness of Census}
\label{sec:comp}

\begin{figure}
\epsscale{0.75}
\plotone{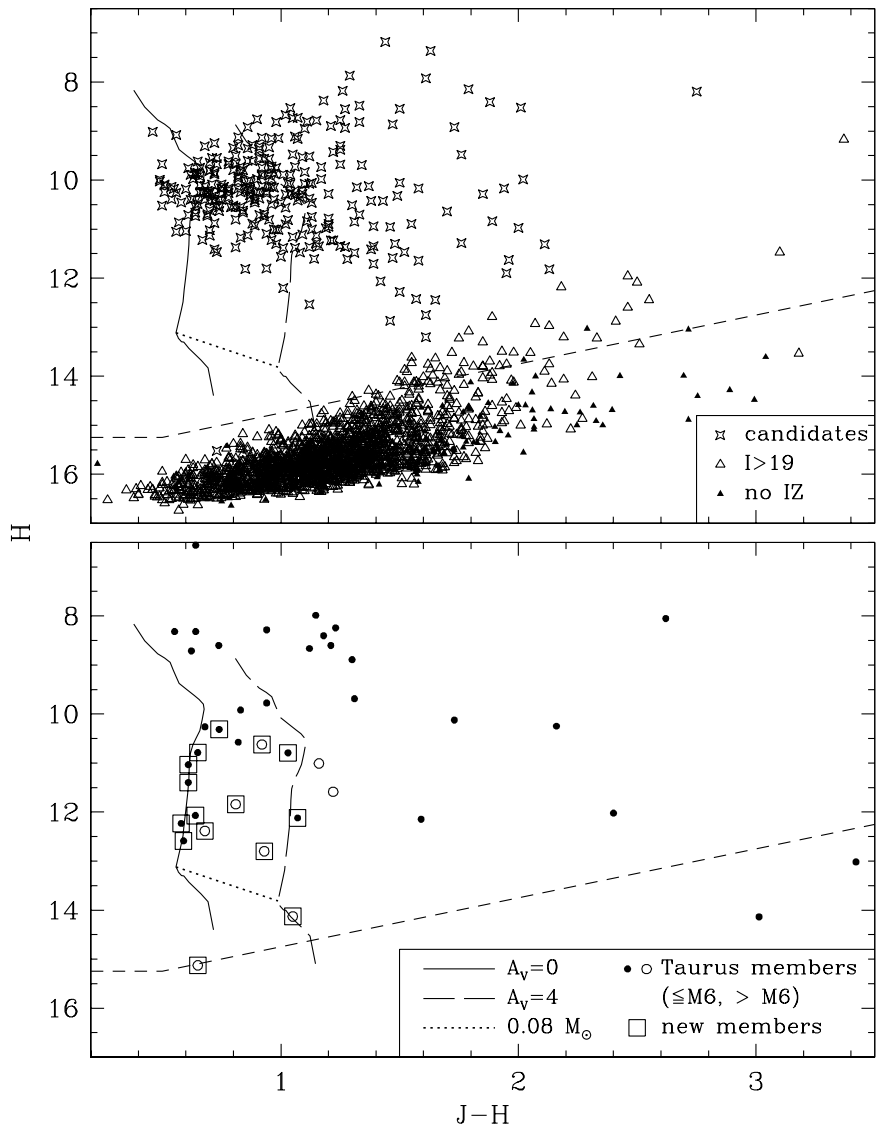}
\caption{
$J-H$ versus $H$ from 2MASS for Field 2 (Figure~\ref{fig:map2}).
The symbols are the same as in Figure~\ref{fig:4sh2}, with the addition of
stars detected only in these IR data ({\it solid triangles}). 
I have omitted the field stars identified through spectroscopy
(Tables~\ref{tab:field}) as well as objects that are probable field stars
by their locations in Figure~\ref{fig:4sh2}.
The 10~Myr isochrone (1.4-0.02~$M_{\odot}$) from the evolutionary models of 
\citet{bar98} is shown for $A_V=0$ ({\it solid line}) and $A_V=4$ ({\it long
dashed line}). The hydrogen burning mass limit at this age and range of 
extinctions is also indicated ({\it dotted line}).
These 2MASS measurements have completeness limits of $J=15.75$ and $H=15.25$ 
({\it short dashed line}).}
\label{fig:jh}
\end{figure}

To evaluate the completeness of my $Iz\arcmin$ survey for new members in
Field 2 in Figure~\ref{fig:map2},
I consider the diagram of $J-H$ versus $H$ in Figure~\ref{fig:jh}.
The completeness limits of the 2MASS photometry are taken to be the magnitudes
at which the logarithm of the number of sources as a function of magnitude
departs from a linear slope and begins to turn over
($J\sim15.75$, $H\sim15.25$).
In Figure~\ref{fig:jh}, I have plotted the 10~Myr isochrone from \citet{bar98}
for $A_V=0$ and 4. I selected this isochrone because most members of Taurus 
fall above it on the H-R diagram (\S~\ref{sec:hr}).
As demonstrated in Figure~\ref{fig:jh}, the IR photometric data should be
complete for members with $M/M_{\odot}\geq0.02$ and $A_V\leq4$,
with the exception of close companions.
In Figure~\ref{fig:jh}, I have omitted the field stars identified with 
spectroscopy as well as objects that are probable field stars by their 
location below the boundaries in Figure~\ref{fig:4sh2}.
The remaining IR sources consist of confirmed members in the bottom panel
and objects that lack spectroscopic data in the top panel. The latter are 
divided into candidate members from Figure~\ref{fig:4sh2}, stars with uncertain
optical photometry ($I>19$), and sources detected only in the IR data.
The data in the top panel of Figure~\ref{fig:jh} indicate that the current
census in Field 2 is complete for $M/M_{\odot}=0.02$-0.2 
and $A_V\leq4$. The upper limit of this mass range is a reflection of
the fact that only candidates with dereddened $I-K_s>2.5$ ($\gtrsim$M4) were 
included in the spectroscopic sample (\S~\ref{sec:cand}). 
At colors bluer than this threshold, which corresponds to $M/M_{\odot}>0.2$
and types of $<$M4, there remains a large number of candidate members and
thus the current census is incomplete based on this survey alone.
Completeness at these higher masses is determined by 
the previous wide-field searches for young stars in Taurus through signatures 
of youth such as emission in H$\alpha$ and X-rays (\citet{ss94,bri99},
references therein), which quote completeness limits of $\sim$M4. However,
because of the difficulties in precisely quantifying completeness in those
surveys, that completeness limit is only a rough estimate, and the applicable
range of extinctions is undetermined. In summary,
the combination of those previous surveys with 
the deeper search in this work should provide a fairly complete census of
Taurus members at $A_V\leq4$ and at most masses above $M/M_{\odot}\geq0.02$,
although some incompleteness may exist at $M/M_{\odot}=0.3$-0.6 (M2-M4).
Just as this survey is incomplete for low-mass members with high
extinction ($A_V>4$), it is not sensitive to objects that are seen
in scattered light (e.g., edge-on disks).

As mentioned in \S~\ref{sec:cand}, I have included in this study the 
spectroscopy of an initial set of candidates from a 225~deg$^2$ survey of 
Taurus, from which seven new members have been found. Because this survey has 
not finished, I do not evaluate its completeness here. However, the new
members produced thus far do have implications for the completeness of the 
previous surveys by \citet{bri98}, \citet{luh00a}, and \citet{bri02}. 
Two of the new members from the 225~deg$^2$ search are within the areas 
considered in those studies (\S~\ref{sec:spatial}), and yet were not found 
in that work.
These two sources were missed because they are slightly brighter than the 
saturation limits of $I\sim13$ in images of \citet{bri98}, \citet{luh00a}, and 
\citet{bri02}. Meanwhile, with spectral types of M4.75 and M6.25, these
objects are just below the completeness limit of $\sim$M4 in the 
older wide-field surveys discussed earlier in this section.
Thus, for a narrow range of spectral types (M4-M6), the census of members 
from \citet{bri98}, \citet{luh00a}, and \citet{bri02} was incomplete. 
In comparison, the 225~deg$^2$ survey is sensitive to members at M4-M9, 
and so it uncovered two members at M4-M6 missed in those studies. 
The presence of only two undiscovered members indicates that the 
incompleteness was minor in the previous surveys. 
The new $Iz\arcmin$ survey in this work does not suffer from
incompleteness at M4-M6 since its saturation limits 
are brighter ($I\sim11.5$), which allows the detection and identification 
of all candidates in this range of types.

\section{The Taurus Stellar Population}
\label{sec:pop}

For all known members of Taurus within Field 1,
photometry, spectral types, extinctions, effective temperatures, and 
bolometric luminosities were compiled by \citet{bri02} and \citet{luh03a}.
Those data were then used to construct an H-R diagram, estimate individual
masses with evolutionary models, and generate an IMF from an extinction-limited
sample. In this section, I repeat these steps for Field 2 from
this work and present a combined IMF for Fields 1 and 2.
The new members found outside of these fields
from the 225~deg$^2$ survey are also included in this analysis up to the 
point of placing them on the H-R diagram, but they will not be added to the IMF
as that survey is not finished and thus lacks well-defined completeness
(\S~\ref{sec:comp}). 

\subsection{H-R Diagram}
\label{sec:hr}

I have estimated
extinctions, effective temperatures, and bolometric luminosities from
the spectral types and photometry for all known members of Taurus in Field 2
and for the additional new members found in the 225~deg$^2$ survey.
These data are listed in Tables~\ref{tab:new} and \ref{tab:old} for
the new and previously known members, respectively.
I have employed the procedures described by \citet{bri02} with the
following modification. \citet{bri02} measured extinctions from optical
spectral only for late-type sources that lacked optical photometry. 
In this work, I have adopted the extinction implied by the optical spectra
for all members in my spectroscopic sample.
Following \citet{luh04}, this reddening is quantified by the 
color excess between 0.6 and 0.9~\micron, denoted as $E(0.6-0.9)$ in 
Figures~\ref{fig:spec1} and \ref{fig:spec2}. These excesses have been
converted to $A_J$ with the relation $A_J=E(0.6-0.9)/1.5$, which is derived
from the extinction laws of \citet{car89} and \citet{rl85} for 
$R_V\equiv A_V/E(B-V)=3.1$ as in \citet{luh04}.
The combined uncertainties in $A_J$, $J$, and BC$_J$ ($\sigma\sim0.14$, 0.03,
0.1) correspond to errors of $\pm0.07$ in the relative values of
log~$L_{\rm bol}$.
When an uncertainty in the distance modulus is included ($\sigma\sim0.2$),
the total uncertainties are $\pm0.11$.
The previously known members contain six pairs with separations less than 
$2\arcsec$. As in my previous work in Taurus, each of these pairs is treated
as one source and appears as a single entry in Table~\ref{tab:old},
resulting in a total of 29 sources in this list. Because seven of these
members lack spectral classifications, extinctions, temperatures, 
and luminosities are not estimated for them.

\begin{figure}
\epsscale{0.8}
\plotone{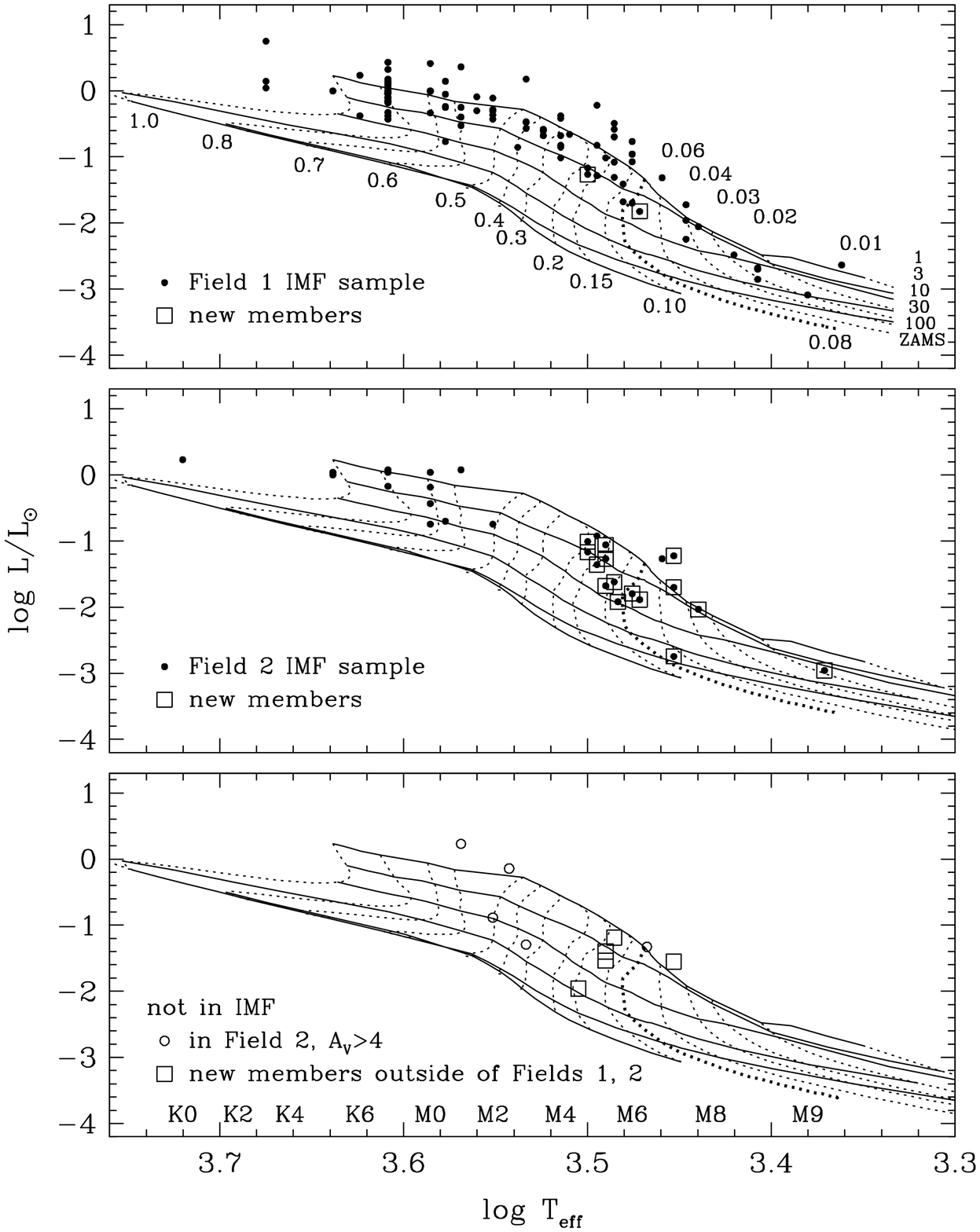}
\caption{\footnotesize
H-R diagram for different samples of Taurus members. 
The IMF from \citet{luh03a} corresponds to an
extinction-limited sample ($A_V\leq4$) for Field 1 from
\citet{luh00a}, \citet{bri02}, and \citet{luh03a} ({\it top}).
The two new members found in this work in Field 1 are shown with that sample.
Taurus members in Field 2 from this work and within this extinction limit 
({\it middle}) have been combined with the upper sample to produce the IMF in
Figure~\ref{fig:imf} ({\it middle}).
Young sources in Field 2 that are beyond the extinction threshold 
of $A_V=4$ and new members from the 225~deg$^2$ survey of all of Taurus
that are outside of Fields 1 and 2 are not included in the IMF ({\it bottom}).
These data are shown with the theoretical evolutionary models of
\citet{bar98} ($0.1<M/M_\odot\leq1$) and \citet{cha00} ($M/M_\odot\leq0.1$),
where the mass tracks ({\it dotted lines}) and isochrones ({\it solid lines}) 
are labeled in units of $M_\odot$ and Myr, respectively. 
}
\label{fig:hr}
\end{figure}

The 44 members in Tables~\ref{tab:new} and \ref{tab:old} with temperature
and luminosity estimates have been placed on the H-R diagram in
Figure~\ref{fig:hr}. Sources that will be included and excluded from the
IMF in \S~\ref{sec:imf} are plotted in the middle and bottom panels,
respectively. For reference, the top panel of Figure~\ref{fig:hr}
shows the known Taurus members that have $A_V\leq4$ and that are within 
Field 1, which comprise the IMF sample from \citet{luh03a}.
As in that previous IMF sample, most of the Taurus members in Field 2
appear above the 10~Myr isochrone of \citet{bar98} and \citet{cha00}.
However, whereas the previous sample has a median age of 1-1.5~Myr,
the members in Field 2 seem to have older ages. When members in
the northern and southern areas in Field 2 are plotted separately, the 
southern sources have the same median age as the previous IMF sample
while the northern members exhibit a median age of 3~Myr. For instance,
all of the members in Field 2 that are above the 1~Myr isochrone
are in the southern area. The southern region is adjacent to some of the 
previously surveyed areas, so it is not surprising that their ages agree. 
Meanwhile, the northern area is far from the other fields in question
and on the edge of the star-forming region. Thus, the members in that field 
could indeed have a different age than the bulk of Taurus. 
Two of the new members, 2MASS~J04163048+3037053 and J04381486+2611399,
appear near the 30~Myr isochrone and thus are anomalously faint for their 
spectral types. Similar sources have
been found in previous surveys of Taurus, IC~348, Orion, and Chamaeleon
\citep{bri02,luh03a,luh03b,sle04,luh04}. The apparent subluminous nature
of these sources may be the result of occultation by circumstellar structures,
which allows only scattered light to reach the observer.
One of these sources, 2MASS~J04381486+2611399, is probably a brown dwarf 
according to its spectral type of M7.25, and exhibits forbidden line emission 
and a large $K$-band excess (\S~\ref{sec:disks}). 
Other examples of subluminous late-type objects are LS-RCrA~1 and KPNO-Tau~12 
\citep{fer01,luh03a}.

\subsection{The Initial Mass Function}
\label{sec:imf}

I now measure the IMF for Fields 1 and 2.
I begin with the IMF for Field 1 from \citet{luh03a}, which consisted of 
the 92 known members in that area with extinctions of $A_V\leq4$. 
The two new members found in this work that were missed in that survey
(\S~\ref{sec:comp}) have $A_V\leq4$ and thus are added to that sample.
As discussed in \S~\ref{sec:comp}, the census of members in Fields 1 and 2
should be complete within this extinction limit
for $M/M_{\odot}\geq0.02$, except for possible incompleteness at 
0.3-0.6~$M_\odot$.
Therefore, I select the known Taurus members within an extinction limit of
$A_V\leq4$ in Field 2 and add them to the IMF from Field 1.
Among the seven members in Table~\ref{tab:old} that lack spectral 
classifications, five sources are not included in the IMF because their
near-IR colors are indicative of $A_V>4$. The other two members are CIDA-7
and HV~Tau~C. CIDA-7 was not placed on the H-R diagram because of uncertainty
in the M2-M3 classification by \citet{bri99}.
Since the colors of CIDA-7 indicate an extinction of $A_V<4$, it should be
included in the IMF. Therefore, I adopt a spectral type of M2.5, which 
corresponds to a mass of $\sim0.5$~$M_\odot$ with the evolutionary models
of \citet{bar98}. Because HV~Tau~C has an edge-on disk \citep{mb00}, it is
omitted from the IMF sample for the reasons described by \citet{bri02}.
After applying these criteria, I arrive at a sample of 33 members in 
Field 2 which are added to the sample of 94 sources in Field 1.
Histograms of the combined IMF sample are shown in terms 
of spectral types and masses in Figures~\ref{fig:histo} and \ref{fig:imf}.
The masses for these objects have been inferred from
the theoretical evolutionary models of \citet{bar98} and \citet{cha00}
for $M/M_\odot\leq1$ and the models of \citet{pal99} for $M/M_\odot>1$ because 
they provide the best agreement with observational constraints \citep{luh03b}. 

Following \citet{bri02} and \citet{luh03b}, I quantify the relative numbers of 
brown dwarfs and stars and the relative numbers of low-mass and high-mass stars
with these ratios:

$$ {\mathcal R}_{1} = N(0.02\leq M/M_\odot\leq0.08)/N(0.08<M/M_\odot\leq10)$$

$$ {\mathcal R}_{2} = N(1<M/M_\odot\leq10)/N(0.15<M/M_\odot\leq1)$$

\noindent
The new Taurus IMF exhibits ${\mathcal R}_{1} = 19/106 = 0.18\pm0.04$ and
${\mathcal R}_{2} = 10/77 = 0.13\pm0.04$, which are higher than, but 
statistically consistent with, the values of $0.14\pm0.04$ and $0.08\pm0.04$ 
for Field 1 \citep{luh03b}.
Because of the potential incompleteness at masses of 0.3-0.6~$M_\odot$
(\S~\ref{sec:comp}), these measurements of ${\mathcal R}_{1}$ and
${\mathcal R}_{2}$ could be overestimated.

\section{Implications of New Taurus IMF}
\label{sec:imfcompare}

Measurements of mass functions 
in Taurus and other star-forming regions can be used to test 
the predictions of models for the formation of stars and brown dwarfs. 
In this section, I first summarize recent observational work in 
star-forming regions and then examine the implications of these data for 
theoretical models. 

\subsection{Recent IMF Measurements}

\subsubsection{Brown Dwarfs}

As reviewed in \S~\ref{sec:intro}, recent spectroscopic surveys of Taurus 
\citep{bri02,luh03a} and IC~348 \citep{luh03b} have reported brown dwarf
fractions ${\mathcal R}_{1}$ that are a factor of two lower than the values
derived from luminosity function modeling in the Trapezium Cluster in Orion 
\citep{luh00b,hc00,mue02}. However, upon spectroscopy of a large sample of
brown dwarf candidates in the Trapezium, \citet{sle04} found a population 
of faint objects with stellar masses, possibly seen in scattered light,
which had contaminated previous photometric IMF samples 
and resulted in overestimates of the brown dwarf fraction in
this cluster. After \citet{sle04} corrected for this contamination,
${\mathcal R}_{1}$ in the Trapezium was a factor of $\sim1.8$ higher 
than the value in Taurus from \citet{luh03a}. This difference would 
be reduced further ($\sim1.4$) if the new measurement of ${\mathcal R}_{1}$ in 
this work is adopted. However, as discussed in the previous section, 
the estimates of ${\mathcal R}_{1}$ in Taurus could be overestimates as well
because of possible incompleteness at masses of 0.3-0.6~$M_\odot$.
In summary, according to the best available data, the brown dwarf fractions 
in Taurus and IC~348 are lower than in the Trapezium, but by a factor (1.4-1.8)
that is smaller than that reported in earlier studies.

\subsubsection{Stars}

\begin{figure}
\epsscale{0.75}
\plotone{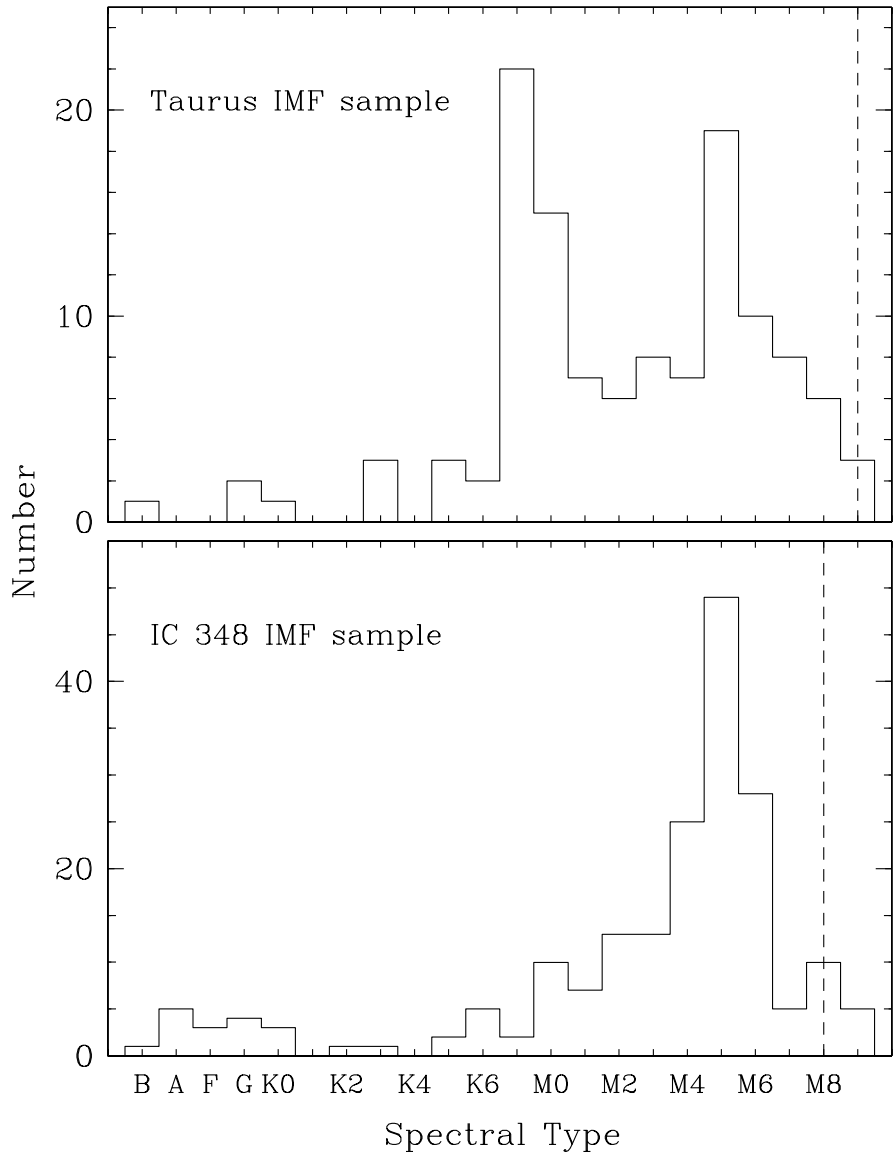}
\caption{
Distributions of spectral types for objects in the IMFs
for Taurus (this work) and for IC~348 \citep{luh03b} that are shown
in Figure~\ref{fig:imf}.
These samples are extinction-limited ($A_V\leq4$) and apply to
Fields 1 and 2 in Taurus (Figure~\ref{fig:map1}) 
and to a $16\arcmin\times14\arcmin$ field in IC~348.
At late types, these samples are nearly 100\% complete for spectral types of 
$\leq$M9 and $\leq$M8, respectively, while the sample for Taurus may be 
incomplete at M2-M4 (\S~\ref{sec:comp}). 
Because the evolutionary tracks for young low-mass stars are mostly
vertical, spectral types should be closely correlated with stellar masses.
As a result, these distributions of spectral types should directly reflect the
IMFs in Taurus and IC~348. 
}
\label{fig:histo}
\end{figure}

\begin{figure}
\epsscale{0.8}
\plotone{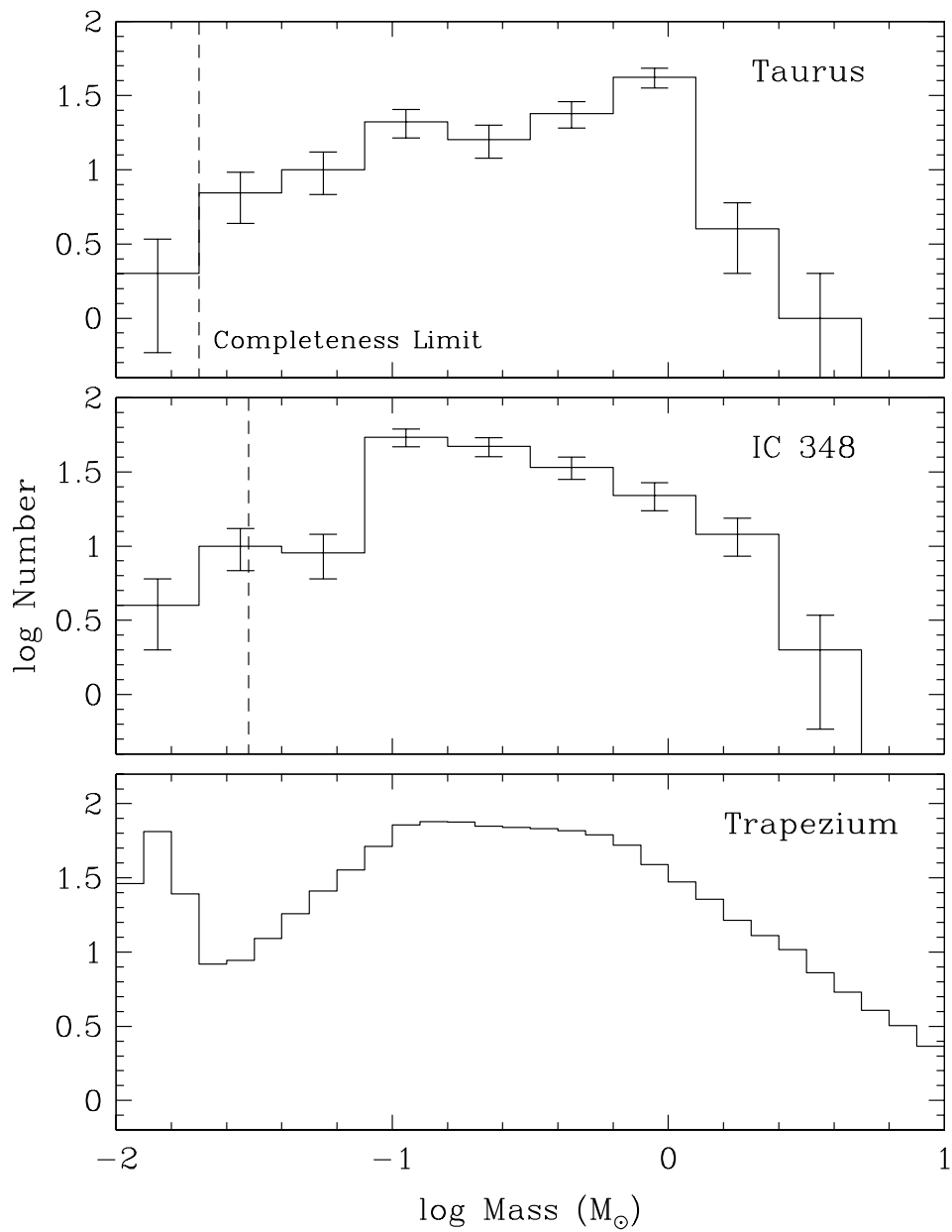}
\caption{
IMFs for extinction-limited samples ($A_V\leq4$) in Fields 1 and 2
in Taurus (this work, {\it top}) and in a 
$16\arcmin\times14\arcmin$ field in IC~348 (\citet{luh03b}, {\it middle}).
These samples are 
unbiased in mass for $M/M_\odot\geq0.02$ and 0.03, respectively, except for 
possible incompleteness at $M/M_\odot=0.3$-0.6 in Taurus (\S~\ref{sec:comp}). 
For comparison, an IMF derived from luminosity function modeling of the 
Trapezium Cluster in Orion is also shown (\citet{mue02}, {\it bottom}).
In the units of this diagram, the Salpeter slope is 1.35.
}
\label{fig:imf}
\end{figure}

In logarithmic units where the Salpeter slope is 1.35, the IMFs for
IC~348 and the Trapezium Cluster rise from high masses down to a solar
mass, rise more slowly down to a maximum at 0.1-0.2~$M_\odot$, and then
decline into the substellar regime \citep{luh03b,mue02,mue03,sle04}. 
In comparison, the IMF for Taurus rises quickly to a peak near 0.8~$M_\odot$ 
and steadily declines to lower masses (\citet{bri02,luh03a}; 
Figure~\ref{fig:imf}).
The distinctive shapes of the IMFs in IC~348 and Taurus are reflected in the
distributions of spectral types, which should comprise good observational 
proxies of the IMFs (\citet{luh03b}; Figure~\ref{fig:histo}).
These significant differences between Taurus and the other two clusters
remain regardless of the degree of incompleteness in the Taurus IMF 
at 0.3-0.6~$M_\odot$. 

At stellar masses, \citet{kro03} found that the IMF for unresolved systems 
from their standard model was nearly identical to the IMF for Taurus from 
\citet{luh03a} and the IMF for the Orion from \citet{mue02}.
As a result, they concluded that these IMFs for Taurus and Orion 
``can be considered very similar if not identical" and ``only in the 
substellar mass regime may the observations indicate different mass functions".
However, this is clearly not the case. As described above,
the IMF for Taurus differs significantly from 
that of IC~348, which in turn closely matches the IMF for the Trapezium
at stellar masses \citep{mue02,mue03,sle04}. This difference between 
Taurus and IC~348 has been shown to be highly statistically significant 
\citep{luh03b}.  As an additional illustration of 
the obvious differences in the IMFs between Taurus and the two other clusters,
the Trapezium IMF from \citet{mue02} is shown with the mass functions of
Taurus and IC~348 in Figure~\ref{fig:imf}. 
The distinct nature of the IMF in Taurus is also evident in the distribution 
of spectral types in Figure~\ref{fig:histo}, as pointed out by \citet{luh03b}.

\subsection{Model Predictions for the IMF}

A simple theoretical explanation for the above variations in the IMF was
offered along with the data by \citet{bri02} and \citet{luh03b}.
These authors suggested that the lower brown dwarf fraction and higher peak
mass in Taurus relative to the Trapezium could reflect differences in 
the typical Jeans masses of the two regions. Other possible sources for these
variations are investigated in this section.

\subsubsection{Ejection}

\citet{kb03b} simulated the brown dwarf fractions in the Trapezium and Taurus
under the assumption that the brown dwarfs form through ejection with the 
same one-dimensional velocity dispersion, $\sigma_{ej}$, in each region. 
This model predicted that ejected 
brown dwarfs remain in the Trapezium for $\sigma_{ej}<6$~km~s$^{-1}$ and escape
a $D=1\arcdeg$ field in Taurus for $\sigma_{ej}>1.5$~km~s$^{-1}$. 
At $\sigma_{ej}\sim2$~km~s$^{-1}$, the simulated brown dwarf fractions 
in the Trapezium and in $1\arcdeg$ Taurus fields agreed
with the measurements available at that time.
As a result, \citet{kb03b} suggested that the lower brown dwarf fraction 
in the Taurus aggregates relative to the Trapezium may be due to the formation 
of brown dwarfs via ejection, and that a distributed population of 
predominantly single brown dwarfs may exist outside of the fields surveyed in 
Taurus. To account for a ``disconcerting discrepancy"
between the brown dwarf fractions in star-forming regions ($\sim0.25$)
and the galactic field ($\sim1$), \citet{kb03b} also examined a variation of 
this model in which the ejection process is not identical in all regions.
In this scenario, brown dwarfs and stars form in roughly equal numbers in
Taurus, but most of the brown dwarfs escape to sufficient distances that
they are missed by current surveys. Meanwhile, brown dwarfs are produced at
a quarter of the rate of stars in the Trapezium, most of which are retained in 
the cluster and thus counted in observations. This model would indicate that 
the brown dwarfs in the field are predominantly formed in low-density regions
like Taurus. In search of further support for the ejection model for brown 
dwarf formation, \citet{kb03b} and \citet{kro03} contended that brown dwarfs 
do not share the same general formation history of stars because their 
binary properties are not a natural extension of those of low-mass stars.

Two aspects of these hypotheses from \citet{kb03b} and \citet{kro03}
warrant discussion:

First, \citet{kb03b} referred to a deficiency of brown dwarfs in
star-forming regions relative to the field, but this deficiency is not 
supported by the available data. They quoted brown dwarf fractions of 0.25
and 1 for star-forming regions and the field, referencing the various
studies of the Trapezium and the analysis of field data by \citet{cha02}.
However, a comparison of substellar mass functions of
young clusters and the field in terms of power-law slopes is preferable
to a comparison of brown dwarf fractions because the fractions reported for
clusters apply to a brown dwarfs at $M/M_\odot=0.02$-0.08, whereas the field
fraction from \citet{cha02} includes the entire mass range for brown dwarfs.
Furthermore, the brown dwarf fraction for the field depends on the functional
form and the lower limit of the substellar IMF, both of which must be
adopted as they are unconstrained by the field data.
Indeed, \citet{cha03} found that the log-normal IMF
from \citet{cha02} implied more brown dwarfs by a factor of three
than observed in recent field surveys. He offered several plausible 
explanations, and an additional possibility is simply that the log-normal 
functional form used to fit the stellar sources is not appropriate for the 
substellar regime.
When substellar mass functions are instead compared in terms of 
power-law slopes (Salpeter is 1.35),
the latest constraints in the field from \citet{cha02} 
($\alpha\lesssim0$) and \citet{all04} 
($-1.5\lesssim\alpha\lesssim0$) are consistent with the mildly 
negative slopes exhibited by the data for star-forming regions, such
as in Figure~\ref{fig:imf}. 

Second, the conclusion by \citet{kro03} and \citet{kb03b} that 
separate formation mechanism are reflected in distinct binary properties
for stars and brown dwarfs is based on a questionable interpretation of
the published data.
In their Figure~5, \citet{kro03} plotted binary fraction as a function of 
mass as measured in multiplicity surveys of field stars. 
They interpreted the data as indicating one value for stars and a
much lower value for brown dwarfs. However, if those data are allowed 
to speak for themselves, they indicate a declining fraction with primary 
mass among the stars from G (57\%) to K (45\%) to early and mid M (42\%).
This trend is even more pronounced after adding the data for M dwarfs (35\%)
from the 8~pc sample of \citet{rg97}.
\citet{kro03} excluded the data of \citet{rg97} from their analysis
because of incompleteness, referencing \citet{hen97}. In doing so,
\citet{kro03} did not account for the effort expended in the years 
since that study in obtaining a complete census of single and companion stars
in the solar neighborhood \citep{rei03a,rei03b}. These latest surveys indicate
only minor incompleteness ($\lesssim15$\%) in the current 8~pc sample, which
has not changed significantly from the sample considered by \citet{rg97}.
For types later than M8, the observed binary fraction of $\sim15$\% 
\citep{rei01,bou03,bur03,clo03,giz03} can be reasonably 
described as a natural extension of the decreasing binary fraction with primary 
mass evident among field stars, particularly given that those measurements 
of objects later than M8
apply only to separations beyond $\sim1$~AU and thus represent lower limits. 
Indeed, a decrease in binary fraction with primary mass of this kind 
is expected from simple random pairing from the same mass function 
without the presence of different formation mechanisms at high and low masses
\citep{kro93}. Considering the continuous decrease in binary fraction with 
primary mass and the fact that the late-M, L, and T samples consist 
of both low-mass stars and brown dwarfs, there is no evidence among these
data for distinct binary properties among brown dwarfs that would indicate
a formation mechanism separate from that of stars. 

When interpreting the observed IMF of brown dwarfs in Taurus and the Trapezium,
in addition to ejection models, \citet{kro03} also examined a scenario
in which stars and brown dwarfs form in the same manner. 
Through their simulations, they found that if populations at the densities
of Taurus and the Trapezium are initially given the same initial brown dwarf 
fractions, that disruption of binaries in the latter over time could 
produce an excess of free-floating brown dwarfs relative to the low-density 
Taurus aggregates, whose companion brown dwarfs remain bound and undisturbed.
\citet{kro03} therefore suggested that the observed differences in brown dwarf 
fractions between Taurus and the Trapezium may not reflect variations in the 
IMF. This is a distinct possibility, particularly given the reduced
difference in brown dwarf fractions for these two regions from the latest
measurements.
However, the analysis of \citet{kro03} considered only the Trapezium and
Taurus and did not attempt to match observations in other clusters such as 
IC~348, which has a similar observed brown dwarf fraction as Taurus even 
with a stellar density that is higher by two orders of magnitude. 

\subsubsection{Fragmentation}

In an attempt to account for the IMF variations between Taurus and Orion,
\citet{goo04} performed hydrodynamical simulations of the 
collapse and fragmentation of Taurus-like cores. 
The IMF produced by their model of Taurus could be characterized by a
log-normal distribution centered at just under a solar mass plus a flat 
distribution below 0.5~$M_\odot$. Most of the objects
in the former population remained in multiple systems within cores, while
the latter sources were predominantly ejected from the cores.
These ejected objects consisted of both brown dwarfs and stars with
mass-independent velocities of 1-2~km~s$^{-1}$, resulting in similar spatial 
distributions at stellar and substellar masses. Both the IMF and spatial
distribution of stars and brown dwarfs produced by this model are in reasonable
agreement with the measurements for Taurus. 
\citet{goo04} concluded that the high value of the peak mass in Taurus was a 
reflection of the properties of cores in the region, which is essentially the
same as the explanation from \citet{bri02} and \citet{luh03b}. 
Those authors suggested that
the different average Jeans masses between Taurus and denser clusters resulted
in the shift in peak mass, which in the model of \citet{goo04} appears in the
form of the adopted core mass function.

Other recent hydrodynamical simulations have made predictions regarding the
IMF of stars and brown dwarfs as well. 
In a model of the fragmentation of a turbulent cloud, 
\citet{bat03} derived an IMF that was characterized by a 
power law at high masses and a flattening (in logarithmic units) somewhere 
below a solar mass into the substellar regime, which is consistent 
with the observations of star-forming regions.
Variations of the IMF have been explored in the subsequent simulations of 
\citet{del04}, who considered the formation of stellar groups for 
different turbulent velocity fields. They found that the predicted 
substellar IMF was dependent on the adopted turbulent field, but the
stellar IMF was not. Thus, while this model might account for differences in 
the brown dwarf fractions among the Trapezium, IC348, and Taurus,
the predicted invariance of the stellar IMF is not observed, at least between
Taurus and other regions.
In contrast, in the turbulent model of \citet{pn02}, the peak mass of the
IMF does change with the level of turbulence, which is in better agreement with
the data.

\subsubsection{Photoevaporation}

Through measurements of UV excesses with $HST$, \citet{rob04} derived 
accretion rates for a sample of stars in the center of the Trapezium,
which they found to be systematically lower than rates estimated for members
of Taurus. They attributed this difference in accretion rates
to the accelerated evaporation of circumstellar disks that results 
from the ionizing radiation from the Trapezium OB stars. \citet{rob04}
suggested that such suppression of accretion in the Trapezium might be 
responsible for the excess of brown dwarfs in the Trapezium relative to the 
Taurus. This hypothesis remains plausible when IC~348 is considered, which is
another population that lacks photo-ionizing stars and exhibits a lower 
brown dwarf fraction than the Trapezium. 
\citet{kb03b} and \citet{wz04} have provided qualitative theoretical 
support for the viability of this scenario.

\subsubsection{Magnetic Tension}

A new theoretical discussion of a possible origin of the IMF has been
presented by \citet{shu04}. In their model of star formation, magnetic 
tension in cloud cores helps define the final masses of stars and brown dwarfs.
While reviewing the available observational constraints for their predictions, 
they suggested that the observed IMFs in Taurus and denser clusters are the
same, citing \citet{kh95} for the Taurus data. As a result, they did not 
investigate in detail whether their model could account for the variations 
in the peak mass found in newer measurements of Taurus \citep{bri02,luh03a} 
and denser star-forming clusters \citep{hc00,mue02,mue03,luh03b,sle04}.
\citet{shu04} did predict that the peak mass and relative numbers of stars 
and brown dwarfs could change with star-forming conditions, and so it remains
possible that this model could reproduce the observed variations in the IMF
discussed in this work.

\section{Conclusions}

Searches for young low-mass stars and brown dwarfs in Taurus 
have covered a steadily increasing fraction 
of this low-density, dispersed star-forming region 
\citep{ss94,bri98,lr98,luh00a,mar01,bri02,luh03a}. 
I have presented a new installment in this series of surveys, the results 
of which are summarized as follows:

\begin{enumerate}

\item
By combining new $I$ and $z\arcmin$ images with near-IR photometry from 2MASS
for two regions encompassing a total area of 4~deg$^2$,
I have constructed extinction-corrected 
color-magnitude diagrams and used them to select candidate members.
I also have drawn an initial sample of candidates from a survey of
225~deg$^2$ encompassing all of the Taurus star-forming region.
Through spectroscopy of these two sets of candidates, I have established that
22 of these sources are new bonafide members of Taurus with spectral types of
M4.5-M9.25, corresponding to masses of 0.3-0.015~$M_\odot$ according to
the theoretical evolutionary models of \citet{bar98} and \citet{cha00}. 
The mass estimates for seven of these new members are below the hydrogen 
burning mass limit.

\item
Several of the new members exhibit signatures of accretion, outflows, and 
circumstellar material in the form of emission in permitted lines
(H$\alpha$, He~I, Ca~II) and blue continuum, forbidden transitions
([O~I], [S~II]), and the near-IR bands. This sample includes the least
massive objects known to show forbidden line emission, which indicates that
processes such as outflows that are common for young stars also occur 
among young brown dwarfs.

\item
The current census for the 4~deg$^2$ area surveyed in this work 
and the 8.4~deg$^2$
area from \citet{bri98}, \citet{luh00a}, and \citet{bri02} is complete at
$A_V\leq4$ for most masses above $M/M_{\odot}\geq0.02$,
although some incompleteness may exist at $M/M_{\odot}=0.3$-0.6 (M2-M4).

\item
For the known members of Taurus within the new 4~deg$^2$ area,
I have estimated extinctions, effective temperatures, and 
bolometric luminosities and have used these results to place the members
on the H-R diagram and estimate individual masses with evolutionary models.
I have constructed an IMF from an extinction-limited sample of known members 
($A_V\leq4$) in the area totaling 12.4~deg$^2$ that has been
surveyed here and in the previous studies cited above.
This IMF peaks at a higher mass (0.8~$M_\odot$) than the mass 
functions in IC~348 and Orion (0.1-0.2~$M_\odot$), which is in agreement
with the results from the earlier surveys of Taurus. 
Because of a slightly higher brown dwarf fraction in the latest IMF 
for Taurus and a lower brown dwarf fraction in the new spectroscopic IMF 
for the Trapezium from \citet{sle04},
Taurus now appears less deficient in brown dwarfs relative to the Trapezium
than previously reported ($\times1.4$-1.8 versus $\times2$). Meanwhile, 
the brown dwarf fraction in Taurus is similar to that observed for IC~348.

\item
The spatial distribution of the new low-mass stars and brown dwarfs 
closely matches that of the more massive members, 
which is consistent the results of previous surveys of Taurus.

\item
Contrary to statements of \citet{kro03} and \citet{kb03b}, 
published data show no evidence for either 1) 
a difference in the relative numbers of brown dwarfs and stars between
the field and star-forming regions or 2) a discontinuity in the binary 
fraction between stars and brown dwarfs in the field. 

\item
Based on the previous two conclusions, I find no evidence to support the 
hypothesis that brown dwarfs form through ejection. 

\item
Some fragmentation models appear to reproduce the variation in the IMF peak mass
observed between Taurus and other star-forming clusters like the Trapezium 
and IC~348 \citep{pn02}. Meanwhile,
the excess of brown dwarfs in the Trapezium relative to IC~348 and Taurus 
may result from the suppression of accretion by disk photoevaporation 
for sources near the Trapezium's OB stars, as recently proposed by 
\citet{rob04}.

\end{enumerate}

\acknowledgements

I thank Lynne Hillenbrand for communicating results prior to publication.
This work was supported by grant NAG5-11627 from the NASA Long-Term Space 
Astrophysics program.

\clearpage

\begin{deluxetable}{lllll}
\tablewidth{0pt}
\tablecaption{Spectroscopy Observing Log\label{tab:log}}
\tablehead{
\colhead{} &
\colhead{} &
\colhead{} &
\colhead{Grating} &
\colhead{Resolution} \\
\colhead{Night} &
\colhead{Date} &
\colhead{Telescope + Instrument} &
\colhead{(lines~mm$^{-1}$)} &
\colhead{(\AA)} 
}
\startdata
1 & 2003 Dec 14 & MMT + Red Channel & 270 & 10 \\
2 & 2003 Dec 15 & MMT + Red Channel & 270 & 10 \\
3 & 2004 Jan 9 & Magellan I + IMACS & 200 & 8 \\
4 & 2004 Jan 10 & Magellan I + IMACS & 200 & 8 \\
\enddata
\end{deluxetable}

\begin{deluxetable}{llllllll}
\tablewidth{0pt}
\tablecaption{Field Stars \label{tab:field}}
\tablehead{
\colhead{2MASS} &
\colhead{$\alpha$(J2000)\tablenotemark{a}} &
\colhead{$\delta$(J2000)\tablenotemark{a}} &
\colhead{$I$\tablenotemark{b}} &
\colhead{$J-H$\tablenotemark{a}} & \colhead{$H-K_s$\tablenotemark{a}}
& \colhead{$K_s$\tablenotemark{a}} &
\colhead{Night} 
}
\startdata
   J04110699+2836226 &   04 11 06.99 &    28 36 22.6 &      \nodata &    0.72 &    0.39 &    12.13 &   2 \\
   J04135549+2823058 &   04 13 55.50 &    28 23 05.9 &      \nodata &    0.75 &    0.35 &    11.16 &   2 \\
   J04173190+2840295 &   04 17 31.91 &    28 40 29.5 &      \nodata &    0.83 &    0.42 &    11.91 &   2 \\
   J04174047+2809529 &   04 17 40.47 &    28 09 53.0 &   15.47 &    1.36 &    0.57 &    10.67 &   2 \\
   J04184767+2834011 &   04 18 47.67 &    28 34 01.2 &   13.51 &    1.19 &    0.47 &     9.65 &   2 \\
   J04185791+2830520 &   04 18 57.92 &    28 30 52.0 &      \nodata &    1.43 &    0.59 &    10.70 &   1 \\
   J04205343+2913401 &   04 20 53.44 &    29 13 40.1 &      \nodata &    0.65 &    0.37 &    12.60 &   4 \\
   J04215823+2657147 &   04 21 58.23 &    26 57 14.7 &      \nodata &    0.93 &    0.49 &    12.14 &   2 \\
   J04261082+2620442 &   04 26 10.82 &    26 20 44.3 &   13.83 &    0.96 &    0.38 &    10.43 &   2 \\
   J04305971+1804237 &   04 30 59.72 &    18 04 23.7 &      \nodata &    0.95 &    0.38 &    10.28 &   1 \\
   J04314495+2327555 &   04 31 44.96 &    23 27 55.5 &      \nodata &    0.90 &    0.50 &    11.48 &   2 \\
   J04331329+2606298 &   04 33 13.29 &    26 06 29.9 &      \nodata &    0.78 &    0.42 &    11.38 &   2 \\
   J04340640+2609471 &   04 34 06.41 &    26 09 47.1 &   14.27 &    1.27 &    0.48 &    10.08 &   2 \\
   J04344110+2701259 &   04 34 41.10 &    27 01 26.0 &      \nodata &    0.75 &    0.37 &    11.11 &   2 \\
   J04390241+2536203 &   04 39 02.41 &    25 36 20.3 &   12.48 &    0.64 &    0.30 &     9.79 &   1 \\
   J04391868+2537325 &   04 39 18.68 &    25 37 32.5 &   15.54 &    0.90 &    0.44 &    11.77 &   1 \\
   J04392527+2539476 &   04 39 25.27 &    25 39 47.6 &   16.24 &    1.23 &    0.60 &    11.22 &   1 \\
   J04400458+2527313 &   04 40 04.59 &    25 27 31.3 &   17.86 &    1.42 &    0.73 &    12.05 &   4 \\
   J04401534+2522106 &   04 40 15.35 &    25 22 10.7 &   14.77 &    1.12 &    0.57 &    10.13 &   1 \\
   J04402642+2613482 &   04 40 26.42 &    26 13 48.3 &   15.37 &    0.92 &    0.53 &    11.47 &   1 \\
   J04411028+2539342 &   04 41 10.28 &    25 39 34.2 &   14.46 &    0.82 &    0.41 &    11.15 &   1 \\
   J04411955+2536174 &   04 41 19.55 &    25 36 17.5 &   18.88 &    0.72 &    0.74 &    14.80 &   4 \\
   J04422817+2524357 &   04 42 28.18 &    25 24 35.7 &   14.46 &    0.79 &    0.41 &    11.18 &   1 \\
   J04422824+2535246 &   04 42 28.24 &    25 35 24.7 &   14.34 &    0.74 &    0.46 &    11.07 &   1 \\
   J04425686+2610546 &   04 42 56.87 &    26 10 54.7 &   13.69 &    0.58 &    0.32 &    11.12 &   1 \\
   J04425863+2515295 &   04 42 58.64 &    25 15 29.6 &   13.70 &    0.67 &    0.34 &    10.86 &   1 \\
   J04425891+2519061 &   04 42 58.91 &    25 19 06.2 &   13.91 &    0.57 &    0.32 &    11.35 &   2 \\
\enddata
\tablenotetext{a}{2MASS Point Source Catalog.}
\tablenotetext{b}{This work and \citet{bri02}.}
\end{deluxetable}

\begin{deluxetable}{lllllllllllllll}
\tabletypesize{\scriptsize}
\rotate
\tablewidth{0pt}
\tablecaption{New Members of Taurus \label{tab:new}}
\tablehead{
\colhead{} &
\colhead{} &
\colhead{} &
\colhead{} &
\colhead{Membership} &
\colhead{} &
\colhead{} &
\colhead{} &
\colhead{} &
\colhead{} &
\colhead{} &
\colhead{} &
\colhead{} &
\colhead{} &
\colhead{} \\
\colhead{2MASS} &
\colhead{$\alpha$(J2000)\tablenotemark{a}} & 
\colhead{$\delta$(J2000)\tablenotemark{a}} &
\colhead{Spectral Type\tablenotemark{b}} &
\colhead{Evidence\tablenotemark{c}} &
\colhead{$T_{\rm eff}$\tablenotemark{d}} &
\colhead{$A_J$} & \colhead{$L_{\rm bol}$} & 
\colhead{$I$\tablenotemark{e}} & \colhead{$I-z\arcmin$\tablenotemark{e}} &
\colhead{$J-H$\tablenotemark{a}} & \colhead{$H-K_s$\tablenotemark{a}} & 
\colhead{$K_s$\tablenotemark{a}} &
\colhead{In IMF?\tablenotemark{f}} &
\colhead{Night}}
\startdata
             J04141188+2811535 &   04 14 11.88 &    28 11 53.5 &           M6.25$\pm$0.5 & $A_V$,NaK,e &   2962 &    0.28 &   0.015 &      \nodata &       \nodata &     0.83 &     0.69 &    11.64  & yes &   2 \\
             J04161210+2756385 &   04 16 12.10 &    27 56 38.6 &          M4.75 &   $A_V$,NaK &   3161 &    0.56 &   0.054 &      \nodata &       \nodata &     1.16 &     0.78 &    10.34  & yes &   2 \\
             J04163048+3037053 &   04 16 30.49 &    30 37 05.3 &           M4.5 &      NaK &   3198 &    0.21 &   0.011 &      \nodata &       \nodata &     0.65 &     0.35 &    12.62  &  no &   4 \\
             J04202555+2700355 &   04 20 25.55 &    27 00 35.5 &          M5.25 & $A_V$,NaK,e &   3091 &    0.56 &   0.029 &      \nodata &       \nodata &     0.86 &     0.50 &    11.51  &  no &   2 \\
             J04213459+2701388 &   04 21 34.60 &    27 01 38.9 &           M5.5 &   $A_V$,NaK &   3058 &    0.49 &   0.065 &      \nodata &       \nodata &     0.93 &     0.53 &    10.44  &  no &   2 \\
             J04284263+2714039 &   04 28 42.63 &    27 14 03.9 &          M5.25 &      NaK &   3091 &    0.14 &   0.040 &      \nodata &       \nodata &     1.04 &     0.61 &    10.46  &  no &   2 \\
             J04380083+2558572$\tablenotemark{g}$ &   04 38 00.84 &    25 58 57.2 &          M7.25 &      NaK &   2838 &    0.17 &   0.060 &   14.71 &     1.27 &     0.92 &     0.53 &    10.10  & yes &   1 \\
             J04381486+2611399 &   04 38 14.86 &    26 11 39.9 &          M7.25 &   NaK,e &   2838 &    0.00 &  0.0018 &   17.84 &     1.09 &     1.05 &     1.15 &    12.98  & yes &   1 \\
             J04390396+2544264 &   04 39 03.96 &    25 44 26.4 &          M7.25 &   NaK,e &   2838 &    0.07 &   0.020 &   15.63 &     1.22 &     0.81 &     0.47 &    11.37  & yes &   1 \\
             J04403979+2519061 &   04 40 39.79 &    25 19 06.1 &          M5.25 &   $A_V$,NaK &   3091 &    0.70 &   0.087 &   14.80 &     1.05 &     1.03 &     0.55 &    10.24  & yes &   1 \\
             J04411078+2555116$\tablenotemark{h}$ &   04 41 10.79 &    25 55 11.7 &           M5.5 & $A_V$,NaK,e &   3058 &    0.70 &   0.024 &   16.24 &     0.99 &     1.07 &     0.67 &    11.45  & yes &   1 \\
             J04414825+2534304 &   04 41 48.25 &    25 34 30.5 &          M7.75 &   NaK,e &   2752 &    0.28 &  0.0092 &   17.03 &     1.24 &     0.93 &     0.58 &    12.22  & yes &   1 \\
             J04442713+2512164$\tablenotemark{i}$ &   04 44 27.13 &    25 12 16.4 &          M7.25 &   NaK,e &   2838 &    0.00 &   0.028 &      \nodata &       \nodata &     0.84 &     0.60 &    10.76  &  no &   2 \\
             J04552333+3027366 &   04 55 23.33 &    30 27 36.6 &          M6.25 &      NaK &   2962 &    0.00 &   0.013 &   15.55 &     1.02 &     0.68 &     0.42 &    11.97  & yes &   2 \\
             J04554046+3039057 &   04 55 40.46 &    30 39 05.7 &          M5.25 &      NaK &   3091 &    0.07 &   0.021 &   14.71 &     0.76 &     0.64 &     0.30 &    11.77  & yes &   1 \\
             J04554535+3019389 &   04 55 45.35 &    30 19 38.9 &          M4.75 &      NaK &   3161 &    0.00 &   0.068 &   13.19 &     0.65 &     0.65 &     0.33 &    10.46  & yes &   1 \\
             J04554757+3028077 &   04 55 47.57 &    30 28 07.7 &          M4.75 &      NaK &   3161 &    0.00 &   0.098 &   13.01 &     0.75 &     0.74 &     0.33 &     9.98  & yes &   3 \\
             J04554801+3028050 &   04 55 48.01 &    30 28 05.0 &           M5.6 &      NaK &   3044 &    0.00 &   0.012 &   15.20 &     0.87 &     0.59 &     0.43 &    12.15  & yes &   3 \\
             J04554969+3019400 &   04 55 49.70 &    30 19 40.0 &             M6 &      NaK &   2990 &    0.00 &   0.016 &   15.02 &     0.88 &     0.58 &     0.37 &    11.86  & yes &   1 \\
             J04555288+3006523 &   04 55 52.89 &    30 06 52.3 &          M5.25 &      NaK &   3091 &    0.00 &   0.054 &   13.70 &     0.74 &     0.61 &     0.30 &    10.73  & yes &   1 \\
             J04555636+3049374 &   04 55 56.37 &    30 49 37.5 &             M5 &      NaK &   3125 &    0.10 &   0.044 &   13.85 &     0.69 &     0.61 &     0.30 &    11.09  & yes &   1 \\
             J04574903+3015195 &   04 57 49.03 &    30 15 19.5 &          M9.25 &      NaK &   2350 &    0.00 &  0.0011 &   19.25 &     1.40 &     0.65 &     0.64 &    14.48  & yes &   1 \\
\enddata
\tablenotetext{a}{2MASS Point Source Catalog.}
\tablenotetext{b}{Uncertainties are $\pm0.25$ subclass unless noted otherwise.}
\tablenotetext{c}{Membership in Taurus is indicated by $A_V\gtrsim1$ and
a position above the main sequence for the distance of Taurus (``$A_V$"),
strong emission lines (``e"), or Na~I and K~I strengths intermediate 
between those of dwarfs and giants (``NaK").}
\tablenotetext{d}{Temperature scale from \citet{luh03b}.}
\tablenotetext{e}{This work.}
\tablenotetext{f}{Indicates whether each object is included in the IMF in 
Figure~\ref{fig:imf} (\S~\ref{sec:imf}).}
\tablenotetext{g}{ITG2.}
\tablenotetext{h}{ITG34.}
\tablenotetext{i}{IRAS 04414+2506 from Kenyon et al.(1994).}
\end{deluxetable}

\begin{deluxetable}{llllllllllllllll}
\tabletypesize{\tiny}
\rotate
\tablewidth{0pt}
\tablecaption{Previously Known Members of Taurus within Field 2\label{tab:old}}
\tablehead{
\colhead{ID} &
\colhead{$\alpha$(J2000)\tablenotemark{a}} & 
\colhead{$\delta$(J2000)\tablenotemark{a}} &
\colhead{Spectral Type} &
\colhead{Ref} &
\colhead{Adopt} &
\colhead{$T_{\rm eff}$\tablenotemark{b}} &
\colhead{$A_J$} & \colhead{$L_{\rm bol}$} & 
\colhead{$R-I$} & \colhead{$I$} &
\colhead{Ref} &
\colhead{$J-H$\tablenotemark{a}} & \colhead{$H-K_s$\tablenotemark{a}} & 
\colhead{$K_s$\tablenotemark{a}} &
\colhead{In IMF?\tablenotemark{c}} 
}
\startdata
                       LkCa 14 &   04 36 19.09 &    25 42 59.0 &                   M0 &     1 &     M0 &     3850 &   0.00 &    0.65 &      \nodata &      \nodata &       \nodata &   0.62 &     0.13 &     8.58 &      yes  \\      
                         GM Tau  &   04 38 21.34 &    26 09 13.7 &               C,M6.5 & 1,2 &   M6.5 &     2935 &   1.15 &   0.047 &      \nodata &   15.04 &      3 &   1.22 &     0.95 &    10.63 &       no  \\      
                         DO Tau  &   04 38 28.58 &    26 10 49.4 &                   M0 &     1 &     M0 &     3850 &   0.66 &     1.1 &    1.24 &   11.30 &   1,3 &   1.23 &     0.94 &     7.30 &      yes  \\      
                        HV Tau A+B &   04 38 35.28 &    26 10 38.6 &                   M1 &     1 &     M1 &     3705 &   0.62 &     1.2 &      \nodata &   11.41 &      3 &   0.94 &     0.38 &     7.91 &      yes  \\      
                        HV Tau C &   04 38 35.49  & 26 10 41.5 & \nodata &     \nodata &     \nodata &     \nodata &   \nodata &     \nodata & \nodata &   \nodata &      \nodata &   \nodata &     \nodata &     \nodata &      no  \\
                   IRAS 04361+2547 &   04 39 13.89 &    25 53 20.9 &                   \nodata &     \nodata &     \nodata &       \nodata &     \nodata &      \nodata &      \nodata &      \nodata &       \nodata &   3.42 &     2.30 &    10.72 &       no  \\      
                        GN Tau A+B &   04 39 20.91 &    25 45 02.1 &               C,M2.5 & 1,2 &   M2.5 &     3488 &   1.17 &    0.72 &      \nodata &   12.73 &      3 &   1.30 &     0.83 &     8.06 &       no  \\      
                   IRAS 04365+2535 &   04 39 35.19 &    25 41 44.7 &                   \nodata &     \nodata &     \nodata &       \nodata &     \nodata &      \nodata &      \nodata &      \nodata &       \nodata &     \nodata &     2.92 &    10.84 &       no  \\      
   CFHT-BD-Tau J043947.3+260139 &   04 39 47.48 &    26 01 40.8 &    M7 & 4,5 &     M7 &     2880 &   0.70 &   0.054 &      \nodata &   15.78 &      3 &   1.16 &     0.68 &    10.33 &      yes  \\      
                      IC2087IR &   04 39 55.75 &    25 45 02.0 &                   \nodata &     \nodata &     \nodata &       \nodata &     \nodata &      \nodata &      \nodata &   15.91 &      3 &   2.62 &     1.78 &     6.28 &       no  \\      
                   IRAS 04370+2559 &   04 40 08.00 &    26 05 25.4 &                   \nodata &     \nodata &     \nodata &       \nodata &     \nodata &      \nodata &      \nodata &   17.06 &      3 &   2.16 &     1.38 &     8.87 &       no  \\      
                        JH 223 &   04 40 49.51 &    25 51 19.2 &                   M2 &     1 &     M2 &     3560 &   0.14 &    0.18 &      \nodata &   12.37 &      3 &   0.83 &     0.43 &     9.49 &      yes  \\      
                      Haro 6-32 &   04 41 04.24 &    25 57 56.1 &             M5 &    3 &     M5 &     3125 &   0.17 &    0.12 &      \nodata &   13.01 &      3 &   0.68 &     0.31 &     9.95 &      yes  \\      
                        ITG 33A &   04 41 08.26 &    25 56 07.5 &                   M3 &    6 &     M3 &     3415 &   1.89 &   0.051 &      \nodata &   16.39 &      3 &   1.59 &     1.06 &    11.09 &       no  \\      
                   IRAS 04381+2540 &   04 41 12.68 &    25 46 35.4 &                   \nodata &     \nodata &     \nodata &       \nodata &     \nodata &      \nodata &      \nodata &      \nodata &       \nodata &   3.01 &     2.60 &    11.54 &       no  \\      
                   IRAS 04385+2550 &   04 41 38.82 &    25 56 26.8 &                   M0 &    7 &     M0 &     3850 &   1.13 &    0.18 &    1.59 &   14.50 &   1,3 &   1.73 &     0.92 &     9.20 &      yes  \\      
                  LkHa 332/G2A+B &   04 42 05.49 &    25 22 56.3 &                   K7 &     1 &     K7 &     4060 &   0.99 &     1.1 &     1.5 &   11.99 &      8 &   1.12 &     0.44 &     8.23 &      yes  \\      
                  LkHa 332/G1A+B &   04 42 07.33 &    25 23 03.2 &                   M1 &     1 &     M1 &     3705 &   1.33 &     1.7 &    1.91 &   12.58 &       9 &   1.18 &     0.46 &     7.95 &       no  \\      
                      V955 Tau A+B &   04 42 07.77 &    25 23 11.8 &                K7,K5 &  1,9 &     K5 &     4350 &   0.90 &     1.0 &    1.30 &   13.33 &       9 &   1.21 &     0.66 &     7.94 &      yes  \\      
                        CIDA-7 &   04 42 21.02 &    25 20 34.4 &               M2-M3? &    10 &      ? &       \nodata &     \nodata &      \nodata &      \nodata &   13.55 &      3 &   0.82 &     0.41 &    10.17 &      yes  \\      
                         DP Tau  &   04 42 37.70 &    25 15 37.5 &                 M0.5 &     1 &   M0.5 &     3778 &   0.41 &    0.20 &    1.14 &   12.43 &   1,3 &   1.31 &     0.93 &     8.76 &      yes  \\      
                         GO Tau  &   04 43 03.10 &    25 20 18.8 &                   M0 &     1 &     M0 &     3850 &   0.77 &    0.37 &    1.32 &   12.26 &   1,3 &   0.94 &     0.44 &     9.33 &      yes  \\      
                        UY Aur A+B &   04 51 47.38 &    30 47 13.5 &                   K7 &     1 &     K7 &     4060 &   0.44 &     1.2 &    1.09 &   10.83 &       1 &   1.15 &     0.75 &     7.24 &      yes  \\      
                   IRAS 04489+3042 &   04 52 06.68 &    30 47 17.6 &                   M2 &    7 &     M2 &     3560 &   3.50 &    0.13 &      \nodata &   17.88 &      3 &   2.40 &     1.64 &    10.38 &       no  \\      
                         GM Aur  &   04 55 10.98 &    30 21 59.5 &                   K7 &    11 &     K7 &     4060 &   0.00 &    0.67 &    0.72 &   10.50 &       1 &   0.74 &     0.32 &     8.28 &      yes  \\      
                       LkCa 19 &   04 55 36.96 &    30 17 55.3 &                   K0 &     1 &     K0 &     5250 &   0.24 &     1.7 &    0.57 &    9.68 &       1 &   0.55 &     0.17 &     8.15 &      yes  \\      
                         AB Aur  &   04 55 45.83 &    30 33 04.4 &                   B9 &     1 &     B9 &    10500 &   0.24 &     137 &    0.11 &    6.85 &       1 &   0.87 &     0.83 &     4.23 &      yes  \\      
                         SU Aur  &   04 55 59.38 &    30 34 01.6 &                   G2 &     1 &     G2 &     5860 &   0.21 &     9.9 &    0.52 &    8.10 &       1 &   0.64 &     0.57 &     5.99 &      yes  \\      
                       HBC 427 &   04 56 02.02 &    30 21 03.8 &                K7,K5 & 1,12 &     K5 &     4350 &   0.17 &     1.1 &    0.76 &   10.05 &       1 &   0.64 &     0.19 &     8.13 &      yes  \\      
\enddata
\tablenotetext{a}{2MASS Point Source Catalog except for HV~Tau~C, whose 
coordinates relative to HV~Tau~AB are from \citet{wg01}.}
\tablenotetext{b}{Temperature scale from \citet{sk82} ($\leq$M0) and
\citet{luh03b} ($>$M0).}
\tablenotetext{c}{Indicates whether each object is included in the IMF in 
Figure~\ref{fig:imf} (\S~\ref{sec:imf}).}
\tablerefs{
(1) \citet{kh95};
(2) \citet{wb03};
(3) this work;
(4) \citet{mar01};
(5) \citet{bri02};
(6) \citet{mar00};
(7) \citet{ken98};
(8) \citet{har94};
(9) \citet{wg01};
(10) \citet{bri99};
(11) \citet{har95};
(12) \citet{ste01}.}
\end{deluxetable}

\clearpage

\clearpage

\clearpage

\clearpage

\clearpage

\clearpage

\clearpage

\clearpage

\clearpage

\clearpage

\clearpage

\clearpage

\clearpage

\end{document}